\documentclass[fp]{jpsj3} % (Full Paper) [fp] 
\usepackage{txfonts}
\usepackage[dvipdfmx]{graphicx}
\usepackage{dcolumn}
\usepackage{bm}
\usepackage{xcolor}
\usepackage{amsmath,amssymb}

\newcommand{\Trs}[1]{\Tr_{#1}}

\renewcommand{\vec}[1]{\mbox{\boldmath{$#1$}}}
\newcommand{\bea}{\begin{eqnarray}}
\newcommand{\eea}{\end{eqnarray}}
\newcommand{\beas}{\begin{eqnarray*}}
\newcommand{\eeas}{\end{eqnarray*}}

\newcommand{\calH}{{\cal H}}

\newcommand{\calS}{{\cal S}}
\newcommand{\calF}{{\cal F}}
\newcommand{\calG}{{\cal G}}

\newcommand{\vecf}{\mbox{\boldmath{$f$}}}

\newcommand{\vecs}{\mbox{\boldmath{$s$}}}

\begin{document}

%\preprint{APS/123-QED}

\title{de Almeida-Thouless Stability and Replica Symmetry Breaking in a Hybrid Watermarking Model with Image and Message Decoding}
%% Statistical mechanical evaluation of a spread-spectrum watermarking model with image restoration II\\ AT stability of a hybrid system with  message decoding and  image decoding 

\author{Tatsuya Uezu$^1$\thanks{E-mail: uezu@cc.nara-wu.ac.jp,  retired}, 
Kao Hayashi$^{1}\thanks{Present affiliation: Sony Corporation}$, 
Masaki Kawamura$^2$\thanks{E-mail: kawamura@sci.yamaguchi-u.ac.jp}}

\inst{$^1$ Graduate School of Humanities and Sciences, Nara Women's  University, Kitauoyanishi-machi, Nara 630-8506, Japan,\\%  (retired)\\
  % $^2$ Sony Corporation\\
$^2$ Graduate School of Sciences and Technology for Innovation, Yamaguchi University, Yoshida 1677-1, Yamaguchi 753-8512, Japan}

%\author{Tatsuya Uezu}\email{uezu@cc.nara-wu.ac.jp}
%\affiliation{ Graduate School of Humanities and Sciences, Nara Women's  University, Kitauoyanishi-machi, Nara 630-8506, Japan, (retired)}
%\author{Kao Hayashi}\altaffiliation{Present affiliation: Sony Corporation}\affiliation{ Graduate School of Humanities and Sciences, Nara Women's  University, Kitauoyanishi-machi, Nara 630-8506, Japan}


%\author{Masaki Kawamura}
%\affiliation{Graduate School of Sciences and Technology for Innovation, Yamaguchi University, Yoshida 1677-1, Yamaguchi 753-8512, Japan}
%\affiliation{
%\affiliation{%


%\date{\today}

\abst{
  We investigate the de Almeida-Thouless (AT) stability of the replica symmetric (RS) solution for a spread-spectrum
  watermarking model incorporating Bayes-estimation-based image restoration. In our previously proposed
  model,\cite{Kawamura.et.al2019} which simultaneously performs message decoding and image restoration under Gaussian
  channel noise, a discrepancy between the theoretical predictions of the RS solution and Markov chain Monte Carlo
  (MCMC) simulations was observed in specific parameter regimes. In this paper, by analyzing the blind case where both the original
  message and image are unknown,  we explicitly derive the conditions for the AT stability of the RS solution
 by clarifying the structure of the replicon eigenspace for the hybrid system. Our
  results
  reveal that AT stability is broken precisely in the parameter region where the RS solution deviates from
  MCMC simulations, thereby successfully explaining the  mechanism underlying this discrepancy
  in terms of replica symmetry breaking (RSB).
}

%\keywords{Suggested keywords}%Use showkeys class option if keyword display desired
\maketitle
%\tableofcontents

%%%%%%%%%%%%%%%%%%%%%%%%%%%%%%%%%%%%%%%%%%%%%%%
\section{INTRODUCTION}
\label{sec:intro}

The inferential restoration of high-dimensional signals from degraded observations remains a central paradigm across statistical physics, information theory, and signal processing. In structured estimation tasks, the Bayesian framework provides an optimal methodology by incorporating prior knowledge of the target signal. In particular, systems exhibiting long-range interactions---such as dense neural networks, multi-user communication
networks, and spin glasses (e.g., the Sherrington-Kirkpatrick model \cite{SK.1975})---offer powerful mathematical models for complex priors.
For instance, a statistical-mechanical formulation has shown that image restoration and error-correcting decoding achieve optimal performance at the Nishimori temperature, a result illustrated by the exact solution of the infinite-range model, \cite{NishimoriWong1999} and has since been extended to watermarking models and joint restoration schemes using parity-check codes to clarify the static and dynamic properties of inter-signal interactions.\cite{Kawamura.et.al2019,Kawaguchi2021,Kawaguchi2021_b}

In our previous work,\cite{Kawamura.et.al2019} we proposed a spread-spectrum watermarking model integrated with image restoration based on Bayesian estimation. In this framework, we introduced several prior probabilities to model the target signals and adopted a Gaussian channel model to represent attacks from unauthorized users. Conceptually, this architecture is a hybrid model that combines the performance evaluation of code-division multiple-access (CDMA) multi-user detectors used in mobile communications with the assessment of image reconstruction capabilities.
Utilizing the replica method --- a technique from statistical mechanics --- and  drawing upon the seminal work by Tanaka on CDMA multi-user detection,\cite{TTanaka2002} we derived the free energy under the replica symmetry (RS) ansatz, along with the saddle-point equations (SPEs) and the expressions for the bit error rates (BERs) of both the embedded message and the host image.
%%%
Through Markov chain Monte Carlo (MCMC) numerical simulations, we confirmed that the theoretical RS predictions generally agreed well with the numerical results. Furthermore, we investigated both the informed case (where the original host image is known) and the blind case (where both the original message and image are unknown), finding that the performance difference between these two scenarios remained small provided the embedding rate and attack intensity were kept low.
%%% 
However, a noticeable discrepancy between the RS theoretical predictions and MCMC simulation results persisted within specific parameter regions in our previous study.\cite{Kawamura.et.al2019} This deviation strongly indicates a breakdown of the RS assumption. To rigorously analyze the domain of validity for the RS solution, it is necessary to evaluate its stability against fluctuations in replica space, a criterion known as the de Almeida-Thouless (AT) stability. \cite{AT1978}

In this paper, to clarify the mechanism underlying the breakdown of the RS ansatz
in this hybrid system, we derive the expression for the free energy without assuming replica symmetry \cite{K.Hayashi2019} and evaluate the AT stability of the RS solution.
The main contributions of this work are threefold:
(1) We present a concise and self-contained formulation of the hybrid watermarking and image restoration model
 with slight updates in notation compared to our previous work~\cite{Kawamura.et.al2019}),
(2) we perform a comprehensive AT stability analysis by clarifying the structure of the replicon
  eigenspace and explicitly evaluating the replicon modes of the Hessian matrix, thereby determining the parameter regime where the RS ansatz breaks down, and
(3) we validate the stability boundaries against extensive MCMC simulations, demonstrating how the onset of replica symmetry breaking (RSB) accounts for the previously observed theoretical discrepancies.

The structure of this  paper is as follows. Section~\ref{sec:formulation} formulates the present model and Bayesian inference framework in a self-contained manner. Section~\ref{sec:replica_rs} presents the derivation of the RS solution using the replica trick. Section~\ref{sec:at_stability}
details the AT stability analysis and the derivation of the replicon mode eigenvalues.
Section~\ref{sec:numerical_results} compares the analytical predictions with MCMC simulations.
Finally, Sect.~\ref{sec:conclusion} provides concluding remarks.
Mathematical details and explicit calculations are relegated to the Appendices.

%%%%%%%%%%%%%%%%%%%%%%%%%%%%%%%%%%%%%%%
\section{FORMULATION}
\label{sec:formulation}
In this section, we formulate the proposed model in a self-contained manner. 
The original host image is represented as an $N$-bit vector $\vec{f}=(f_1, f_2, \dots, f_N)^{\top}$, 
where $f_\mu \in \{-1, +1\}$ for $\mu=1, \dots, N$, and $(\cdot)^{\top}$ denotes the transpose operation. We embed a $K$-bit message $\vec{s}=\left(s_1, s_2, \dots, s_K\right)^{\top}$ with $s_k \in \{-1, +1\}$ as follows. Each message bit $s_k$ is spread across all $N$ pixels using a specific pseudo-random spreading code $\vec{\xi}_k=\left(\xi_k^{1}, \xi_k^{2}, \dots, \xi_k^{N}\right)^{\top}$, where $\xi_k^\mu \in \{-1, +1\}$. The watermark $\vec{w}=(w_1, w_2, \dots, w_N)^{\top}$ is formed as the linear combination of the $K$ spread messages. Specifically, the watermark embedded in the $\mu$-th pixel is given by
\begin{equation}
w_{\mu} = \frac{1}{\sqrt{K}} \sum_{k=1}^{K} \xi^{\mu}_k s_k, \quad \mu=1,2,\dots, N.
\end{equation}
The stego image $\vec{X}=(X_1, X_2, \dots, X_N)^{\top}$ is constructed by superimposing the watermark $\vec{w}$ onto the original image $\vec{f}$, i.e., $\vec{X} = \vec{f} + \vec{w}$. 
%%%
During transmission, the stego image is degraded by external disturbances (such as malicious attacks or channel impairments), which we model as additive white Gaussian noise (AWGN). Consequently, the tampered image $\widetilde{\vec{f}}=\left(\widetilde{f}_1, \widetilde{f}_2, \dots, \widetilde{f}_N\right)^{\top}$ is expressed as
\begin{eqnarray}
  \widetilde{f}_{\mu} &=& f_{\mu} + w_{\mu} + n_{\mu} \nonumber \\
  &=& f_{\mu} + \frac{1}{\sqrt{K}}\sum_{k=1}^K \xi_k^{\mu}s_k + n_{\mu}, \quad \mu=1,2, \dots, N,
\end{eqnarray}
where $n_{\mu}$ represents the noise sampled from a Gaussian distribution ${\cal N}(0, \sigma_0^2)$ with mean zero and variance $\sigma_0^2$. The conditional likelihood of observing $\widetilde{\vec{f}}$ given the true signals and channel noise level is therefore
\begin{eqnarray}
 P\left(\widetilde{\vec{f}} \;\middle|\; \vec{s}, \vec{f}, \sigma_0, \xi \right) 
 &=& \prod_{\mu=1}^{N} P_s\left({\widetilde{f}_{\mu}} \;\middle|\; \vec{s}, \vec{f}, \sigma_0, \xi \right) \nonumber \\
 &=& \left(2\pi\sigma_0^2\right)^{-\frac{N}{2}} \exp\left[-\frac{1}{2\sigma_0^2} \sum_{\mu=1}^N \left({\widetilde{f}_{\mu}} - \frac{1}{\sqrt{K}}\sum_{k=1}^K \xi_k^{\mu}s_k - f_{\mu} \right)^2 \right],
\end{eqnarray}
where $\xi$ collectively denotes the set of all spreading codes $\{\xi_k^\mu\}$.

To perform a statistical-mechanical analysis, we make the following assumptions regarding the prior probabilities.
First, for the message vector $\vec{s}$, assuming no prior knowledge, we adopt an uninformative uniform prior:
\begin{eqnarray}
 P\left(\vec{s}\right) &=& \frac{1}{2^K}.
\end{eqnarray}
For the host image, we employ an infinite-range Ising model to capture long-range spatial correlations:
\begin{eqnarray}
P\left(\vec{f} \;\middle|\; \alpha_0\right) &=& \frac{1}{Z_f }\exp\left[\frac{\alpha_0}{2N} \left(\sum_{\mu=1}^{N}f_{\mu}\right)^2\right], \\
Z_f &=& \Trs{\vec{f}} \exp\left[\frac{\alpha_0}{2N} \left(\sum_{\mu=1}^{N}f_{\mu}\right)^2\right],
\end{eqnarray}
where $\Trs{\vec{f}}$ denotes the summation over all possible configurations of $f_\mu = \pm 1$ ($\mu=1,\dots,N$). More generally, in what follows, $\Trs{\vec{a}}$ represents the trace (configuration sum) over all components of a vector $\vec{a}$.
A ferromagnetic phase emerges when $\alpha_0 > 1$, which constitutes the main focus of our study. By Bayes' theorem, the true posterior probability distribution is given by
\begin{eqnarray}
  P\left(\vec{s},\vec{f} \;\middle|\; {\widetilde{\vec{f}}}, \sigma_0, \alpha_0, \xi \right)
& = & \frac{1}{Z_0} P \left( {\widetilde{\vec{f}}} \;\middle|\; \vec{s},\vec{f}, \sigma_0, \xi \right) P\left(\vec{s}\right) P\left(\vec{f} \;\middle|\; \alpha_0 \right), \\
Z_0 &=& \Trs{\vec{s}} \Trs{\vec{f}} P\left( {\widetilde{\vec{f}}} \;\middle|\; \vec{s},\vec{f}, \sigma_0, \xi \right) P\left(\vec{s} \right) P\left(\vec{f} \;\middle|\; \alpha_0 \right).
\end{eqnarray}
%%%
In practical inference, however, the true data-generating parameters $\sigma_0$ and $\alpha_0$ are typically unknown to the receiver. To model this hyperparameter mismatch, we introduce assumed parameters $\sigma$ and $\alpha$ inside the estimator. Furthermore, we consider the blind estimation scenario where both the original message $\vec{s}$ and image $\vec{f}$ are unknown. Let $\vec{x}=(x_1, x_2,\dots,x_K)^{\top}$ and $\vec{g}=(g_1, g_2, \dots,g_N)^{\top}$ be the candidate variables for $\vec{s}$ and $\vec{f}$, respectively. The assumed posterior probability distribution realized by the estimator is written as
\begin{eqnarray}
 P\left(\vec{x},\vec{g}\;\middle|\; {\widetilde{\vec{f}}}, \sigma, \alpha, \xi\right) 
 &=& \frac{1}{Z} P\left({\widetilde{\vec{f}}} \;\middle|\; \vec{x},\vec{g}, \sigma, \xi \right) P\left(\vec{x}\right) P\left(\vec{g}\;\middle|\; \alpha\right), \\
 Z &=& \Trs{\vec{x}} \Trs{\vec{g}} P\left({\widetilde{\vec{f}}}\;\middle|\; \vec{x},\vec{g}, \sigma, \xi \right) P\left(\vec{x}\right) P\left(\vec{g}\;\middle|\; \alpha\right).
\end{eqnarray}
%%%
The signal reconstruction is executed using Maximum A Posteriori Marginal (MPM) estimation. Concatenating the system state into $\vec{S} = \left(s_1, \dots, s_K, f_1, \dots, f_N \right)^{\top}$ and its estimate into $\widehat{\vec{S}}=(\widehat{s}_1,\dots,\widehat{s}_K,\widehat{f}_1,\dots,\widehat{f}_N)^{\top}$, the $i$-th component $\widehat{S}_i$ is estimated by maximizing the marginal posterior probability:
\begin{equation}
\widehat{S}_i = \mathop{\mathrm{argmax}}_{Y_i} \sum_{(\vec{x}, \vec{g}) \backslash Y_i} P\left(\vec{x}, \vec{g} \;\middle|\; {\widetilde{\vec{f}}}, \sigma, \alpha, \xi \right),
\end{equation}
where $Y_i$ corresponds to the $i$-th element of $(\vec{x},\vec{g})$, and $\sum_{(\vec{x},\vec{g}) \backslash Y_i}$ represents the summation over all variables in $(\vec{x}, \vec{g})$ except $Y_i$.

To quantify the typical performance, we evaluate the average over
all realization uncertainties in the system, including messages $\vecs$, host images $\vecf$,
spreading codes $\xi$, and tampered observations $\widetilde{\vecf}$. In other words, these variables are considered to be quenched variables.
The configuration average with respect to the quenched variables of an arbitrary quantity $A$ is defined by

\begin{eqnarray}
 \left[A\right]_{\mathrm{av}} &\equiv& \frac{1}{2^K Z_f} \int \left[\prod_{\mu=1}^{N}d{\widetilde{f}_{\mu}} \right] \Trs{\vec{s}} \Trs{\vec{f}} \left\langle P\left({\widetilde{\vec{f}}} \;\middle|\; \vec{s},\vec{f}, \sigma_0, \xi \right) P\left(\vec{s}\right) P\left(\vec{f} \;\middle|\; \alpha_0\right) A \right\rangle_{\xi},
 \label{eq:config_av}
\end{eqnarray}
where $\left\langle \cdot \right\rangle_{\xi}$ denotes the average over the spreading codes $\xi_k^\mu$, which are assumed to be independently and identically distributed with uniform probability over $\{-1, +1\}$. 
%%%
The typical properties of the quenched system are governed by the averaged log-partition function $[\ln Z]_{\mathrm{av}}$:
\begin{eqnarray}
 \left[ \ln Z\right]_{\mathrm{av}} 
 &=& \frac{1}{2^K Z_f} \int \prod_{\mu=1}^{N}d{\widetilde{f}_{\mu}} \Trs{\vec{s}} \Trs{\vec{f}} 
 \Biggl\langle P\left({\widetilde{\vec{f}}} \;\middle|\; \vec{s},\vec{f}, \sigma_0, \xi \right) P\left(\vec{s}\right) P\left(\vec{f}\;\middle|\; \alpha_0\right) \nonumber \\
 &\times& \ln \left( \Trs{\vec{x}}\Trs{\vec{g}} P\left({\widetilde{\vec{f}}} \;\middle|\; \vec{x},\vec{g}, \sigma, \xi \right) P\left(\vec{x}\right)P\left(\vec{g}\;\middle|\; \alpha \right) \right) \Biggr\rangle_{\xi}.
\end{eqnarray}
We evaluate $[\ln Z]_{\mathrm{av}}$ using the standard replica trick:
\begin{equation}
\left[\ln Z\right]_{\mathrm{av}} = \lim_{n \to 0}\frac{\left[Z^n\right]_{\mathrm{av}} - 1}{n}.
\end{equation}
Here, the partition function $Z$ is explicitly written as
\begin{eqnarray}
 Z &=& \frac{1}{2^K Z_g} \Trs{\vec{x}} \Trs{\vec{g}} \left(2\pi\sigma^2\right)^{-\frac{N}{2}} \exp\left[-\frac{1}{2\sigma^2}\sum_{\mu=1}^N \left({\widetilde{f}_{\mu}}-\frac{1}{\sqrt{K}}\sum_{k=1}^{K} \xi^{\mu}_k x_k -g_{\mu} \right)^2 \right. \nonumber \\ && \left.
 +\frac{\alpha}{2N} \left(\sum_{\mu=1}^{N}g_{\mu} \right)^2 \right],  \\
Z_g &=& \Trs{\vec{g}} \exp\left[\frac{\alpha}{2N} \left(\sum_{\mu=1}^{N}g_{\mu} \right)^2\right].   
\end{eqnarray}
Introducing $n$ replicas labeled by $a=1, 2, \dots, n$, the replicated partition function averaged over the quenched variables becomes
\begin{eqnarray}
 \left[Z^n\right]_{\mathrm{av}} &=& \frac{1}{2^K Z_f} \int\prod_{\mu=1}^{N}d{\widetilde{f}_{\mu}} \Trs{\vec{s}} \Trs{\left\{\vec{x}^a\right\}} \Trs{\vec{f}} \Trs{\left\{\vec{g}^a\right\}} \left< \left(2\pi\sigma_0^2\right)^{-\frac{N}{2}} \right. \nonumber \\ 
 &\times& \exp\left[-\frac{1}{2\sigma_0^2}\sum_{\mu=1}^N \left({\widetilde{f}_{\mu}} -\frac{1}{\sqrt{K}}\sum_{k=1}^{K} \xi^{\mu}_ks_k -f_{\mu} \right)^2 +\frac{\alpha_0}{2N} \left(\sum_{\mu=1}^{N}f_{\mu} \right)^2 \right] \nonumber \\ 
 &\times& \left. \exp\left[-\frac{1}{2\sigma^2}\sum_{a=1}^n \sum_{\mu=1}^N \left({\widetilde{f}_{\mu}} -\frac{1}{\sqrt{K}} \sum_{k=1}^{K} \xi^{\mu}_k x_k^{a}-g_{\mu}^{a}\right)^2 +\frac{\alpha}{2N}\sum_{a=1}^n \left(\sum_{\mu=1}^{N}g_{\mu}^a \right)^2 \right] \right>_{\xi}, \nonumber \\
 \label{eq:npowerofZ}
\end{eqnarray}
where $\vec{x}^a=(x_1^a, \dots, x_K^a)^{\top}$ and $\vec{g}^a=(g_1^a, \dots, g_N^a)^{\top}$. The symbols $\Trs{\left\{\vec{x}^a\right\}}$ and $\Trs{\left\{\vec{g}^a\right\}}$ represent summations over all replica components. The multiplicative constant $((2\pi\sigma^2)^{N/2} 2^K Z_g)^{-n}$ is irrelevant and is omitted for brevity.

We introduce the order parameters $m_a$ and $q_{ab}$ for the messages, as well as auxiliary variables:
\begin{equation}
m_{a} = \frac{1}{K} \sum_{i=1}^K s_i x_i^a, \quad q_{ab} = \frac{1}{K} \sum_{i=1}^K x_i^a x_i^b,
\end{equation}
\begin{equation}
v_0^\mu = \frac{1}{\sqrt{K}}\sum_{i=1}^K \xi_i^\mu s_i, \quad v_a^\mu = \frac{1}{\sqrt{K}}\sum_{i=1}^K \xi_i^\mu x_i^a.
\end{equation}
Similarly, the order parameters for the images are defined as
\begin{equation}
R_a = \frac{1}{N}\sum_{\mu=1}^N f_{\mu}g_{\mu}^a, \quad r_0 = \frac{1}{N}\sum_{\mu=1}^N f_{\mu}, \quad r_a = \frac{1}{N}\sum_{\mu=1}^N g_{\mu}^a, \quad Q_{ab} = \frac{1}{N}\sum_{\mu=1}^N g_{\mu}^a g_{\mu}^b.
\label{eq:opimage}
\end{equation}
We also define the corresponding conjugate variables denoted with a hat ($\widehat{\,\cdot\,}$).
In the present model, since the integrand factorizes site by site in the thermodynamic limit ($N, K \to \infty$), the averaged partition function can be expressed as follows:
\begin{eqnarray}
 \left[Z^n\right]_{\mathrm{av}} &=& \int d \vec{V} e^{NG}, \label{eq:Zn_integral} \\
 G &=& G_1 + G_2 + \widetilde{G}_3 - \beta \ln 2 - \frac{1}{N}\ln Z_f, \label{eq:G_total}
\end{eqnarray}
where $\beta \equiv K/N$ is the embedding rate, and the integration measure is defined as
\begin{eqnarray}
 d \vec{V} &\equiv& \prod_{a<b} \left[\frac{K i d \widehat{q}_{ab}}{2\pi} dq_{ab} \right] \prod_a \left[\frac{iK d \widehat{m}_a}{2\pi} dm_a \right] d \vec{X}, \\
 d \vec{X} &\equiv& \prod_a \left[\frac{iN dR_ad\widehat{R}_a}{2\pi}\right] \left[\frac{iN dr_0d\widehat{r}_0}{2\pi}\right] \prod_a \left[\frac{i N dr_ad\widehat{r}_a}{2\pi}\right] \prod_{a<b} \left[\frac{i N dQ_{ab}d\widehat{Q}_{ab}}{2\pi}\right].
\end{eqnarray}

The individual components of $G$ are evaluated as follows:
\begin{eqnarray}
G_1 &=& -\beta\sum_{a<b}\widehat{q}_{ab}q_{ab} -\beta\sum_{a}\widehat{m}_{a}m_a, \\
G_2 &=& \beta \ln \left( \Trs{s} \Trs{\left\{x^a\right\}} \exp\left[ \sum_{a<b}\widehat{q}_{ab} x^a x^b + \sum_a \widehat{m}_a s x^a \right] \right),
\end{eqnarray}
where $\mathrm{Tr}_s$ and $\mathrm{Tr}_{\{x^a\}}$ denote the summation over $s \in \{-1, +1\}$ and that over $x^a \in \{-1, +1\}$ for $a = 1, \dots, n$, respectively.
\begin{eqnarray}
e^{NG_3} &=& \int d \vec{X} e^{N \widetilde{G}_3}, \label{eq:g31} \\
e^{\widetilde{G}_3} &=& \Trs{f} \Trs{\{g^a\}} \frac{\sigma^{n+1}}{\sqrt{\sigma^2 + n(1+ \sigma_0^2)}} \int D t \left[ \prod_{a} \int \frac{d\widehat{v}_a}{\sqrt{2\pi}} \right] \exp\left[{\cal L} + G_4\right], \label{eq:g32} \\
G_4 &=& -\left(\widehat{r}_0 r_0 + \sum_a \widehat{r}_a r_a + \sum_a \widehat{R}_a R_a + \sum_{a<b} \widehat{Q}_{ab} Q_{ab} \right) + \frac{1}{2} \left( \alpha_0 r_0^2 + \alpha \sum_a r_a^2 \right) \nonumber \\
&& + \widehat{r}_0 f + \sum_a \widehat{r}_a g^a + \sum_a \widehat{R}_a f g^a + \sum_{a<b} \widehat{Q}_{ab} g^a g^b, \label{eq:g41} \\
{\cal L} &=& \frac{\sigma^2}{2} n( \xi t)^2 - \frac{\sigma^2}{2} \sum_a \left(\widehat{v}_a\right)^2 - i n \xi t U \frac{\sigma^2}{\sigma^2 + n(1+ \sigma_0^2)} + i \sigma^2 \xi t \sum_a \widehat{v}_a \nonumber \\
&& + U \frac{\sigma^2}{\sigma^2 + n(1+ \sigma_0^2)} \sum_a \widehat{v}_a - i \sum_a \widehat{v}_a g^a - \frac{1}{2}\sum_{a}(\widehat{v}_a)^2 - \sum_{a<b} q_{ab} \widehat{v}_a \widehat{v}_b + {\cal O}\left(n^2\right), \nonumber \\ \label{eq:l51} \\
U &=& \sum_{a} m_{a} \widehat{v}_a + if, \label{eq:U_def} \\
\xi &=& \frac{1}{\sigma}\sqrt{ \frac{1+\sigma_0^2}{\sigma^2 + n(1+\sigma_0^2)} }. \label{eq:xi_def}
\end{eqnarray}
where $D t \equiv e^{-t^2 / 2} dt / \sqrt{2 \pi}$.
Detailed derivations for these expressions are provided in Appendix A.

In the thermodynamic limit $N \to \infty$, the multi-dimensional integral over the order parameters
and their conjugates is evaluated via the saddle-point method:
\begin{equation}
\int d \vec{V} e^{NG} \sim e^{N \mathop{\mathrm{Extr}} G},
\end{equation}
where $\mathop{\mathrm{Extr}} G$ represents the extremum of $G$ with respect to the order parameters and their conjugates. Consequently,
the averaged ``free energy'' per pixel is defined as $F=\frac{1}{N} \left[\ln Z\right]_{\mathrm{av}}$, and in the limit $n\to0$ , it is obtained as follows:
\begin{equation}
F = \lim_{n \to 0} \frac{G}{n}.
\end{equation}

%%%%%%%%%%%%%%%%%%%%%%%%%%%%%%%%%%%%%%%%%%%%%%
\section{RS SOLUTION}
\label{sec:replica_rs}

Assuming replica symmetry, the RS solution and its analytical properties were derived in our previous study.\cite{Kawamura.et.al2019} In this section, we present the explicit form of the RS free energy and the corresponding SPEs. For completeness, the detailed derivation of the free energy $F_{\mathrm{RS}} = \lim_{n \to 0} \frac{G_{\mathrm{RS}}}{n}$ is outlined in Appendix~\ref{sec:RS_solution}.

Under the RS ansatz, the free energy per pixel $F_{\mathrm{RS}} = G_{\mathrm{RS}}^1$ is expressed as
\begin{eqnarray}
G_{\mathrm{RS}}^1
&=& \frac{1}{2}\beta q \widehat{q} - \beta m \widehat{m} - \beta \frac{\widehat{q}}{2} + \beta \int Dz \ln \left[ 2 \cosh \left( \sqrt{\widehat{q}} z + \widehat{m} \right) \right] \nonumber \\
&& - \frac{(1-Q)\left(2m - q - (1+\sigma_0^2)\right)}{2(\sigma^2 + 1 - q)^2} - \frac{1 - R}{\sigma^2 + 1 - q} - \widehat{r} r - \widehat{R} R - \frac{1}{2} \widehat{Q}(1-Q) \nonumber \\
&& + \frac{1}{2}\alpha r^2 - \frac{1}{2} \ln \left( \frac{\sigma^2 + 1 - q}{\sigma^2} \right) + \frac{1}{2} \frac{2m - q - (1+\sigma_0^2)}{\sigma^2 + 1 - q} \nonumber \\
&& + \frac{1}{2 \cosh \widehat{r}_0} \mathrm{Tr}_{f} e^{\widehat{r}_0 f} \int Ds \ln \left[ 2\cosh \Xi(s, f) \right],
\end{eqnarray}
where $Dz \equiv e^{-z^2/2} dz$, and $\Xi(s, f) \equiv \widehat{r} + \widehat{R} f + \sqrt{\widehat{Q}} s$.
%%% 
Extremizing $G_{\mathrm{RS}}^1$ with respect to the order parameters and their conjugate variables yields the following set of SPEs for the RS solution: %self-consistent 
\begin{eqnarray}
\widehat{r}_0 &=& \alpha_0 r_0, \quad r_0 = \tanh \widehat{r}_0, \label{eq:spe_r0} \\
\widehat{r} &=& \alpha r, \quad r = \frac{1}{2 \cosh \widehat{r}_0} \mathrm{Tr}_{f} e^{\widehat{r}_0 f} \int Ds \tanh \left( \widehat{r} + \widehat{R} f + \sqrt{\widehat{Q}} s \right), \label{eq:spe_r} \\
\widehat{m} &=& \frac{1}{\beta(\sigma^2 + 1 - q)} - \frac{1-Q}{\beta(\sigma^2 + 1 - q)^2}, \label{eq:spe_mhat} \\
m &=& \int Du \tanh \left( \sqrt{\widehat{q}} u + \widehat{m} \right), \label{eq:spe_m} \\
\widehat{q} &=& \frac{1}{\beta (\sigma^2 + 1 - q)^2} \left[ q - 2m + \sigma_0^2 + 2(1-R) + Q \right] \nonumber \\
&& - \frac{2}{\beta(\sigma^2 + 1 - q)^3} (1-Q)(q - 2m + 1 + \sigma_0^2), \label{eq:spe_qhat} \\
q &=& \int Du \tanh^2 \left( \sqrt{\widehat{q}} u + \widehat{m} \right), \label{eq:spe_q} \\
\widehat{R} &=& \frac{1}{\sigma^2 + 1 - q}, \label{eq:spe_Rhat} \\
R &=& \frac{1}{2 \cosh \widehat{r}_0} \mathrm{Tr}_{f} f e^{\widehat{r}_0 f} \int Ds \tanh \left( \widehat{r} + \widehat{R} f + \sqrt{\widehat{Q}} s \right), \label{eq:spe_R} \\
\widehat{Q} &=& \frac{q - 2m + 1 + \sigma_0^2}{(\sigma^2 + 1 - q)^2}, \label{eq:spe_Qhat} \\
Q &=& \frac{1}{2 \cosh \widehat{r}_0} \mathrm{Tr}_{f} e^{\widehat{r}_0 f} \int Ds \tanh^2 \left( \widehat{r} + \widehat{R} f + \sqrt{\widehat{Q}} s \right). \label{eq:spe_Q}
\end{eqnarray}

%%%%%%%%%%%%%%%%%%%%%%%%%%%%%%%%%%%%%%%%%%%
\section{ AT STABILITY}
\label{sec:at_stability}

The stability of the RS solution is evaluated via the AT stability condition\cite{AT1978},
which is determined by the eigenvalues of the Hessian matrix ${\cal H}_{\mathrm{rp}}$ corresponding to the replicon mode.\cite{MPV1987} 
The matrix ${\cal H}_{\mathrm{rp}}$ has the block structure given by
\begin{equation}
{\cal H}_{\mathrm{rp}} = \begin{pmatrix}
H(P_3, Q_3, R_3) & -\beta E_{n_1} & 0_{n_1,n_1} & H(\widetilde{P}_3, \widetilde{Q}_3,\widetilde{R}_3) \\
-\beta E_{n_1} & H(P'_3, Q'_3, R'_3) & 0_{n_1,n_1} & 0_{n_1,n_1} \\
0_{n_1,n_1} & 0_{n_1,n_1} & 0_{n_1,n_1} & -E_{n_1} \\
H(\widetilde{P}_3, \widetilde{Q}_3,\widetilde{R}_3) & 0_{n_1,n_1} & -E_{n_1} & H(\widehat{P}_3, \widehat{Q}_3, \widehat{R}_3)
\end{pmatrix},
\end{equation}
where $H(P, Q, R)$ is an $n_1 \times n_1$ matrix, $E_{n_1}$ denotes the $n_1 \times n_1$ identity matrix, $0_{n_1,n_1}$ is the $n_1 \times n_1$ zero matrix, and $n_1 \equiv \binom{n}{2} = n(n-1)/2$. The detailed derivations of the matrix elements of ${\cal H}_{\mathrm{rp}}$, such as $P_3$, are provided in Appendix~\ref{sec:app_at_stability}.

Let $H_{(ab)(cd)}$ denote the $((ab), (cd))$ component of $H(P, Q, R)$, where $a, b, c, d$ represent replica indices subject to $a < b$ and $c < d$. The elements $H_{(ab)(cd)}$ are categorized according to the number of shared replica indices:
\begin{equation}
H(P,Q, R)_{(ab)(cd)} = \begin{cases}
P & \text{if } (cd) = (ab), \\
Q & \text{if } c \in \{a,b\} \text{ or } d \in \{a,b\} \text{ (with } (cd) \neq (ab)\text{)}, \\
R & \text{if } \{a,b\} \cap \{c,d\} = \emptyset.
\end{cases}
\label{repliconH}
\end{equation}

\subsection{Eigenvalue problem}
We consider the eigenvalue problem for ${\cal H}_{\mathrm{rp}}$:
\begin{equation}
  {\cal H}_{\mathrm{rp}} \vec{u} = \lambda \vec{u},
  \label{eigenu}
\end{equation}
where $\vec{u}$ is a $4n_1$-dimensional column vector. 
\if0
\begin{equation}
\vec{u} = \begin{pmatrix}
\vec{v} & \vec{0}_{n_1} & \vec{0}_{n_1} & \vec{0}_{n_1} \\
\vec{0}_{n_1} & \vec{v} & \vec{0}_{n_1} & \vec{0}_{n_1} \\
\vec{0}_{n_1} & \vec{0}_{n_1} & \vec{v} & \vec{0}_{n_1} \\
\vec{0}_{n_1} & \vec{0}_{n_1} & \vec{0}_{n_1} & \vec{v}
\end{pmatrix}
\begin{pmatrix}
w_1 \\
w_2 \\
w_3 \\
w_4
\end{pmatrix} = U \begin{pmatrix}
w_1 \\
w_2 \\
w_3 \\
w_4
\end{pmatrix},
\end{equation}
\fi
%We put $\vec{u} = U \vec{w}$,
Based on the form  of ${\cal H}_{\mathrm{rp}}$, we set $\vec{u}=\vec{w}\otimes \vec{v}$, 
where $\vec{w} = (w_1, w_2, w_3, w_4)^{\top}$ is a $4$-dimensional column vector and 
$\vec{v} = (v_1, v_2, \cdots, v_{n_1})^{\top}$ is an $n_1$-dimensional column vector, 
 and $\otimes$ denotes the Kronecker product. We assume that $\vec{w}, \vec{v}$
 and thus $\vec{u}$ are normalized.  Then, $\vec{u}=U\vec{w}$ holds, where
  a $4n_1 \times 4$ matrix $U$ is expressed as
\begin{equation}
  U \equiv E_4 \otimes \vec{v} =
\begin{pmatrix}
\vec{v} & \vec{0}_{n_1} & \vec{0}_{n_1} & \vec{0}_{n_1} \\
\vec{0}_{n_1} & \vec{v} & \vec{0}_{n_1} & \vec{0}_{n_1} \\
\vec{0}_{n_1} & \vec{0}_{n_1} & \vec{v} & \vec{0}_{n_1} \\
\vec{0}_{n_1} & \vec{0}_{n_1} & \vec{0}_{n_1} & \vec{v}
\end{pmatrix},
\end{equation}
where $E_4$ is the $4\times 4$ identity matrix,
$\vec{0}_{n_1}$ is the $n_1$-dimensional zero vector.%, $\vec{v}$ is an $n_1$-dimensional column vector.
%and $\otimes$ denotes the Kronecker product.
%We seek a $4 \times 4$ matrix $C$ satisfying
We introduce a $4 \times 4$ matrix $C$ satisfying
\begin{equation}
{\cal H}_{\mathrm{rp}} U = U C.
\label{eq:for_C}
\end{equation}
Using this relation, Eq.~(\ref{eigenu}) becomes 
\begin{equation}
  C \vec{w} = \lambda \vec{w}.
  \label{eigenw}
\end{equation}
%That is, the problem reduces to the eigenvalue problem for the matrix $C$.
%That is, the eigenvalue problem for ${\cal H}_{\mathrm{rp}} $ , Eq.~\eqref{eigenu} reduces to that  for $C$, Eq. ~\eqref{eigenw}.
That is, the eigenvalue problem for ${\cal H}_{\mathrm{rp}} $ reduces to that  for $C$.
Denoting the $(i,j)$ element of $C$ as $c_{ij}$, Eq.~\eqref{eq:for_C} yields 
\begin{eqnarray}
& H\vec{v} = c_{11}\vec{v}, \quad \widetilde{H}\vec{v} = c_{14}\vec{v}, \quad H' \vec{v} = c_{22}\vec{v}, \quad \widehat{H} \vec{v} = c_{44}\vec{v}, & \\
& c_{12} = c_{21} = -\beta, \quad c_{34} = c_{43} = -1, \quad c_{ij} = 0\; (\text{otherwise}), &
\end{eqnarray}
where $H(P_3, Q_3, R_3)$, $H(P'_3, Q'_3, R'_3)$, $H(\widehat{P}_3, \widehat{Q}_3, \widehat{R}_3)$, and $H(\widetilde{P}_3, \widetilde{Q}_3, \widetilde{R}_3)$ are abbreviated as $H, H', \widehat{H}$, and $\widetilde{H}$, respectively. Thus, $\vec{v}$ must be a simultaneous eigenvector of $H, \widetilde{H}, H'$, and $\widehat{H}$. 
%%%
Each component of $\vec{v}$ is indexed by a pair of replica indices $(ab)$ with $a < b$. Following the standard
 recipe,\cite{MPV1987,Nishimori2001} we analyze three symmetry-distinct classes of eigenvectors: Type I, Type II, and Type III.
 For a Type I vector $\vec{v}_{\mathrm{I}}$, all components take the same value:
\begin{equation}
\vec{v}_{\mathrm{I}} \propto (1, 1, \dots, 1)^{\top}.
\end{equation}
Let $\vec{v}^{(\theta)}$ be a Type II vector specified by a fixed replica index  $\theta \in \{1, 2, \dots, n\}$. Its $(ab)$-component takes $f_2$ if $a = \theta$ or $b = \theta$, and $g_2$ otherwise.
Let  $\vec{v}^{(\theta, \nu)}$ be a Type III vector specified by a pair of distinct indices $(\theta, \nu)$ with $\theta < \nu$. Its  $(\theta\nu)$-component takes $e_3$. If either $a$ or $b$ equals $\theta$ or $\nu$, the $(ab)$-component takes $f_3$, and takes $g_3$ otherwise.
By imposing mutual orthogonality among Type I, Type II, and Type III vectors, we obtain the relations,
\begin{equation}
    f_2 = -\frac{n-2}{2} g_2,\; e_3 = \frac{(n-2)(n-3)}{2} g_3,\;  f_3 = -\frac{n-3}{2} g_3.
\end{equation}
Consequently, in the limit $n \to 0$, the action of $H$ on these eigenvectors yields
\begin{eqnarray}
H \vec{v}_{\mathrm{I}} &=& (P_3 - 4Q_3 + 3R_3) \vec{v}_{\mathrm{I}}, \label{eq:eigen_T1} \\
H \vec{v}^{(\theta)} &=& (P_3 - 4Q_3 + 3R_3) \vec{v}^{(\theta)}, \label{eq:eigen_T2} \\
H \vec{v}^{(\theta, \nu)} &=& (P_3 - 2Q_3 + R_3) \vec{v}^{(\theta, \nu)}. \label{eq:eigen_T3}
\end{eqnarray}
Thus, $P_3 - 4Q_3 + 3R_3$ is the eigenvalue corresponding to Type I and Type II eigenvectors (with degeneracies 1
 and $n-1$, respectively, giving a total degeneracy of $n$), 
 while $P_3 - 2Q_3 + R_3$ corresponds to Type III eigenvectors with degeneracy $n(n-3)/2$.
 The total dimensionality of the eigenspace is $n_1$.\\
% Next, we establish the characteristic equation for $C$,
 % whose eigenvalues coincide with those of ${\cal H}_{\mathrm{rp}}$. The matrix $C$ is written as
We next solve the eigenvalue problem for $C$, Eq. (\ref{eigenw}).
\if0
 \begin{equation}
   C\vec{w}=\lambda \vec{w}.
\end{equation}
$\lambda$ is also the eigenvalue of ${\cal H}_{\mathrm{rp}}$.
Here, $\lambda$ is also an eigenvalue of ${\cal H}_{\mathrm{rp}}$ as shown below.
\begin{equation}
{\cal H}_{\mathrm{rp}} \vec{u}={\cal H}_{\mathrm{rp}} U\vec{w}=UC\vec{w}=\lambda U\vec{w}=\lambda \vec{u}.
\end{equation}
\fi
The matrix $C$ is written as
\begin{equation}
C = \begin{pmatrix}
X & -\beta & 0 & \widetilde{X} \\
-\beta & X' & 0 & 0 \\
0 & 0 & 0 & -1 \\
\widetilde{X} & 0 & -1 & \widehat{X}
\end{pmatrix}.
\end{equation}
For Type I vectors, the parameters in $C$ are
\begin{equation}
X = P_3 - 4 Q_3 + 3R_3, \; 
X' = P'_3 - 4 Q'_3 + 3R'_3, \; 
\widehat{X} = \widehat{P}_3 - 4 \widehat{Q}_3 + 3\widehat{R}_3, \; 
\widetilde{X} = \widetilde{P}_3 - 4 \widetilde{Q}_3 + 3\widetilde{R}_3;
\end{equation}
whereas for Type III vectors, they are given by
\begin{equation}
X = P_3 - 2 Q_3 + R_3, \; 
X' = P'_3 - 2 Q'_3 + R'_3, \; 
\widehat{X} = \widehat{P}_3 - 2 \widehat{Q}_3 + \widehat{R}_3, \; 
\widetilde{X} = \widetilde{P}_3 - 2 \widetilde{Q}_3 + \widetilde{R}_3.
\end{equation}
%%%
The characteristic equation $\det(\lambda E_4 - C) = 0$ leads to the quartic algebraic equation:
\begin{equation}
\lambda^4 + a_3 \lambda^3 + a_2 \lambda^2 + a_1 \lambda + a_0 = 0, \label{eq:4dim_eq}
\end{equation}
with coefficients
\begin{eqnarray}
a_3 &=& -(X + X' + \widehat{X}), \\
a_2 &=& X X' + X \widehat{X} + X' \widehat{X} - \widetilde{X}^2 - \beta^2 - 1, \\
a_1 &=& X + X' - X X' \widehat{X} + \beta^2 \widehat{X} + \widetilde{X}^2 X', \\
a_0 &=& \beta^2 - X X'.
\end{eqnarray}
%%%
For Type I vectors, the explicit expressions for $X, X', \widehat{X}$, and $\widetilde{X}$ are
\begin{eqnarray}
X &=& \kappa^4(1-2\zeta) + 2\kappa^6\left[(2-3\zeta)Q - 2R + 3\zeta\right] \nonumber \\
&& + \kappa^8 \biggl( -1 - 4\zeta + 4\zeta^2 - 2Q\left[1 - 8\zeta(1-\zeta)\right] + 3\overline{R}\left[1 - 4\zeta(1-\zeta)\right] \nonumber \\
&& \qquad + 4(1-2\zeta)R - 4(1-2\zeta)\overline{T} \biggr), \\
X' &=& \beta(1 - 4q + 3\overline{r}), \\
\widehat{X} &=& 1 - 4Q + 3\overline{R}, \\
\widetilde{X} &=& (1 - 4Q + 3\overline{R} - 2\overline{T} + 2R)\kappa^4 - 2 \kappa^2 \widehat{Q}(8Q - 3\overline{R} - 1),
\end{eqnarray}
where we introduced
\begin{equation}
\zeta \equiv \frac{2m - q - (1+\sigma_0^2)}{\sigma^2 + 1 - q}, \quad \kappa^2 \equiv \frac{1}{\sigma^2 + 1 - q},
\end{equation}
\begin{equation}
\overline{r} \equiv \langle x^a x^b x^c x^d \rangle_{2, \mathrm{RS}}, \quad \overline{R} \equiv \langle g^a g^b g^c g^d \rangle_{3, \mathrm{RS}}, \quad \overline{T} \equiv \langle f g^a g^b g^c \rangle_{3, \mathrm{RS}}.
\end{equation}
The statistical averages $\langle \cdot \rangle_{2, \mathrm{RS}}$ and $\langle \cdot \rangle_{3, \mathrm{RS}}$ are explicitly defined in Appendix~\ref{sec:app_at_stability}.
%%%
On the other hand, for Type III vectors, $X, X', \widehat{X}$, and $\widetilde{X}$ are expressed as
\begin{eqnarray}
X &=& \kappa^4 - 2(1-Q) \kappa^6 + (1 - 2Q + \overline{R})\kappa^8, \\
X' &=& \beta(1 - 2q + \overline{r}), \quad \widehat{X} = 1 - 2Q + \overline{R}, \quad \widetilde{X} = (1 - 2Q + \overline{R})\kappa^4.
\end{eqnarray}
As for the dimensionality, four eigenvectors of $C$ correspond to each eigenvector of Types I, II, and III.
$\vec{u}$ is expressed as $\vec{u} = \vec{w} \otimes \vec{v}$, where $\vec{w}$ is an eigenvector of $C$ and $\vec{v}$
is an eigenvector of $H$ (with $H$ taking the form shown in equation (\ref{repliconH})).
Consequently, the relationship between the total replicon eigenspace
$\mathcal{V}_{\mathrm{rp}}^{\mathrm{total}}$, the total eigenspace $\mathbb{R}^4$ associated with $C$,
and the replicon eigenspace $\mathcal{V}_{\mathrm{rp}}$ (spanned by eigenvectors of types I, II, and III) is given as follows:
\begin{equation}
{\cal V}_{\mathrm{rp}}^{\mathrm{total}} =\mathbb{R}^4 \otimes {\cal V}_{\mathrm{rp}}.  
\end{equation}
Therefore, the total dimension is $4n_1$, as expected.

In the next section, we analyze the solutions of Eq.~\eqref{eq:4dim_eq} numerically to determine the conditions for the AT stability of the RS solution and compare the analytical findings with numerical simulation results.

%%%%%%%%%%%%%%%%%%%%%%%%%%%%%%%%%%%%%%%%%%
\section{NUMERICAL RESULTS}
\label{sec:numerical_results}

We evaluated the performance of the RS solution and verified its stability across various system parameters, including the image size $N$, embedding rate $\beta = K/N$, channel noise levels $\sigma_0$ and $\sigma$, and hyperparameter values $\alpha_0$ and $\alpha$. Throughout the numerical calculations and MCMC
simulations, we assumed matched channel noise parameters, i.e., $\sigma = \sigma_0$.
%%% 
In the MCMC  simulations, to suppress finite-size effects while maintaining computational feasibility, system sizes ranging from $N = 1024$ to $10000$ were selected depending on the target parameter settings and numerical resolution requirements. All simulation results were averaged over 200 independent random realizations. The specific value of $N$ used for each dataset is indicated in the corresponding figure captions.\\

\subsection{Determination of the AT stability condition}

 High noise levels in communication channels physically correspond to high-temperature regimes in physical systems. At high temperatures, thermal fluctuations restore replica symmetry, rendering the replica symmetric (RS) solution stable. Consequently, the RS solution is expected to remain AT-stable in high-noise regimes.
  To establish this explicitly, we evaluate the asymptotic behavior of the variables such as order parameters and the parameters $X, X', \widehat{X},$ and $\widetilde{X}$ in the high-noise limit, i.e., $\sigma = \sigma_0 \gg 1$, while keeping $\alpha_0, \alpha,$ and $\beta$ fixed at order $\mathcal{O}(1)$.
From the SPEs, we obtain
\begin{equation}
\widehat{m}, \widehat{q}, m, q, \widehat{R}, \widehat{Q} = \mathcal{O}(\sigma^{-2}), \quad
\overline{r} \simeq 3 \widehat{q}^2 = \mathcal{O}(\sigma^{-4}).
\end{equation}
The remaining variables are of order ${\cal O}(1)$ and are expressed as 
  \begin{align}
  &  r\simeq \tanh \widehat{r},  \quad \widehat{r}=\alpha r, \quad  R \simeq r_0 r,\quad Q\simeq r^2, \\
    &   \overline{R}\simeq r^4, \quad \overline{T}\simeq r_0 r^3 .
  \end{align}
Here, $r_0$ is determined by $r_0 = \tanh(\alpha_0 r_0)$, while $r$ is approximately governed by $r \simeq \tanh(\alpha r)$. Both equations admit non-zero solutions for $\alpha_0 > 1$ and $\alpha > 1$, respectively.
Then, for Type I and Type III vectors, noting that $\kappa^2 \sim \sigma^{-2} \ll 1$, the parameters $X$s are evaluated as
  \begin{align}
  X &= \mathcal{O}(\sigma^{-4}), \\
  X' &\simeq \beta = \mathcal{O}(1), \\
  \widehat{X} &\simeq 
\left\{
\begin{array}{ll}
  (1-r^2)(1-3 r^2)  &\mbox{Type I},\\
  (1 - r^2)^2  &\mbox{Type III},\\
\end{array}
\right.\\
  \widetilde{X} &=
  \mathcal{O}(\sigma^{-4}).
  \end{align}
Retaining the leading-order terms, the characteristic equation simplifies to
\begin{align}
\det(\lambda E_4 - C) = \left( \lambda^2 - \widehat{X} \lambda - 1 \right) \left( \lambda^2 - X' \lambda - \beta^2 \right) = 0,
\end{align}
which yields four real eigenvalues explicitly:
\begin{align}
\widehat{\lambda}_{\pm} = \frac{\widehat{X} \pm \sqrt{\widehat{X}^2 + 4}}{2}, \quad
\lambda'_{\pm} = \frac{X' \pm \sqrt{{X'}^2 + 4\beta^2}}{2}.
\end{align}
Reflecting the saddle-point nature of the RS solution, there are always two positive eigenvalues ($\widehat{\lambda}_{+}, \lambda'_{+}$) and two negative eigenvalues ($\widehat{\lambda}_{-}, \lambda'_{-}$). 
Arranging the eigenvalues in descending order, $\lambda_1 \ge \lambda_2 \ge \lambda_3 \ge \lambda_4$, the AT-stability condition corresponds to maintaining this sign structure without any eigenvalue crossing zero:
\begin{equation}
\lambda_1 \ge \lambda_2 > 0 > \lambda_3 \ge \lambda_4
\label{eq:at_stability_condition}
\end{equation}
for both Type I and Type III modes.
  As can be seen from Eqs. \eqref{eq:eigen_T1} and \eqref{eq:eigen_T2}, the Type II eigenvectors share the same eigenvalue as the Type I eigenvectors, $P_3-4Q_3+3R_3$. Therefore, the AT stability of the Type II mode is already guaranteed by that of the Type I mode.  
We solved the quartic algebraic equation derived from the replicon Hessian matrix $\mathcal{H}_{\mathrm{rp}}$ for both Type I and Type III  modes using the explicit root formulas.   
For both modes, we confirmed the condition in Eq. (\ref{eq:at_stability_condition})
when the noise level is high within the range of parameters examined.
Therefore,
Eq. \eqref{eq:at_stability_condition} defines the AT-stable region for the present hybrid model.
Furthermore, our calculations reveal that the AT stability criterion for Type I
eigenvectors is satisfied throughout the entire parameter range examined,
rendering the Type III mode the primary driver of replica symmetry breaking.
Thus, the AT line for the Type III mode is determined by the condition $a_0 = 0$, which yields
\begin{align}
\beta^2 &= X X'.
\end{align}
This condition can be written explicitly as
%which can be written explicitly as  %that is,
\begin{align}
\beta &= \left( \kappa^4 - 2(1-Q)\kappa^6 + (1 - 2Q + \overline{R})\kappa^8 \right) (1 - 2q + \overline{r}).
\end{align}

\subsection{Matched hyperparameter case: $\alpha = \alpha_0$}

\begin{figure}[tb]
  \centering
  % ----  (a), (b) ----
  \begin{minipage}{0.48\textwidth}
    \centering
    \includegraphics[width=\textwidth]{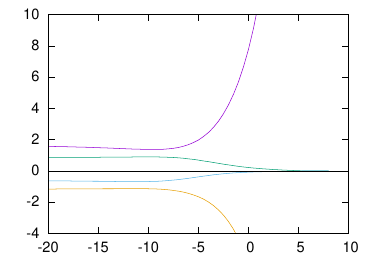}\\
    (a) All $\lambda$ (Type I)\label{fig:sub_1a}
  \end{minipage}%
  \hfill
  \begin{minipage}{0.48\textwidth}
    \centering
    \includegraphics[width=\textwidth]{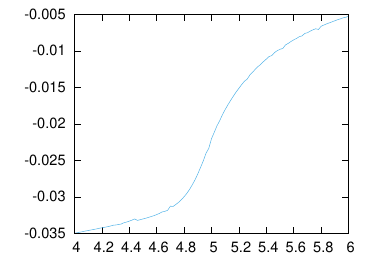}\\
    (b) Enlarged $\lambda_3$ (Type I)\label{fig:sub_1b}
  \end{minipage}

  \vspace{1em} % 

  % ----  (c), (d) ----
  \begin{minipage}{0.48\textwidth}
    \centering
    \includegraphics[width=\textwidth]{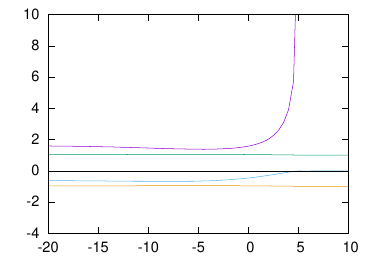}\\
    (c) All $\lambda$ (Type III)\label{fig:sub_1c}
  \end{minipage}%
  \hfill
  \begin{minipage}{0.48\textwidth}
    \centering
    \includegraphics[width=\textwidth]{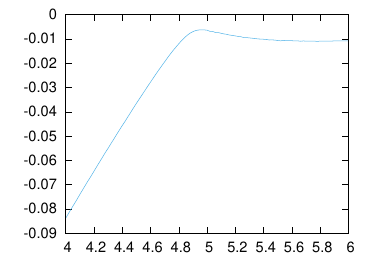}\\
    (d) Enlarged $\lambda_3$ (Type III)\label{fig:sub_1d}
  \end{minipage}
  \caption{(Color online) Eigenvalue behavior for matched hyperparameters $\alpha_0 = 1.5$, $\alpha = 1.5$, and $\beta = 1.0$. %The curves correspond to $\lambda_1$ (purple), $\lambda_2$ (green), $\lambda_3$ (light blue), and $\lambda_4$ (magenta).
(a), (b) Eigenvalues calculated using Type I vectors; (c), (d) eigenvalues calculated using Type III vectors.
    (a), (c) Overview of all eigenvalues.
    From top to bottom, the curves correspond to $\lambda_1$, $\lambda_2$, $\lambda_3$, and $\lambda_4$.
    (b), (d) Enlarged view of $\lambda_3$.}
  \label{fig.1}
\end{figure}

First, we examine the matched hyperparameter scenario where $\alpha = \alpha_0 = 1.5$. 
This parameter-matched
condition corresponds to the Nishimori line, along which the decoding and restoration performance is 
known to be optimal in related infinite-range models.\cite{NishimoriWong1999}
Figure~\ref{fig.1} displays the signal-to-noise ratio dependence of the eigenvalues for $\beta = 1.0$
as a representative example, since similar stability behaviors are observed for $\beta = 0.125$ and $0.5$.
In all figures, the horizontal axis represents the energy-per-bit to noise-power-spectral-density ratio
$E_b/N_0$ defined by
\begin{equation}
\frac{E_b}{N_0} \equiv -10 \log_{10}\left(2\sigma^2\right) \quad [\mathrm{dB}],
\end{equation}
which increases as the channel noise level decreases. The vertical axis denotes the eigenvalues $\lambda_i$ ($i=1,\dots,4$). 
As shown in Fig.~\ref{fig.1}, the RS solution remains AT-stable across the entire range of $E_b/N_0$ under matched
hyperparameters. This is generally consistent with the result that replica symmetry breaking does not occur
on the Nishimori line, both in the original spin-glass context \cite{NishimoriSherrington2001} and,
more specifically, in Bayes-optimal inference problems.\cite{ZdeborovaKrzakala2016}

%% As shown in Fig.~\ref{fig.1}, the RS solution remains AT-stable across the entire range of $E_b/N_0$ under matched hyperparameters.

\begin{figure}[tb]
  \centering
  % ---- (a), (b) ----
  \begin{minipage}{0.48\textwidth}
    \centering
    \includegraphics[width=\textwidth]{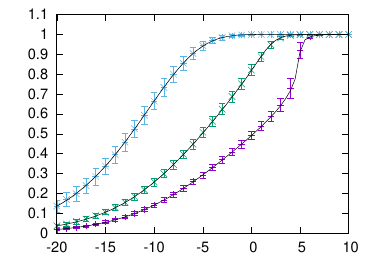}\\
    (a) Order parameter $m$\label{fig:sub_2a}
  \end{minipage}%
  \hfill
  \begin{minipage}{0.48\textwidth}
    \centering
    \includegraphics[width=\textwidth]{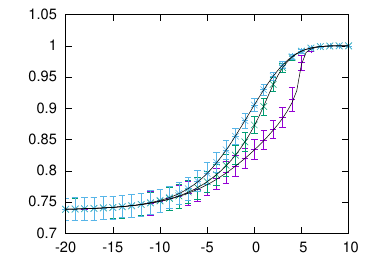}\\
    (b) Order parameter $R$\label{fig:sub_2b}
  \end{minipage}

  \vspace{1em} % 

  % ---- (c), (d) ----
  \begin{minipage}{0.48\textwidth}
    \centering
    \includegraphics[width=\textwidth]{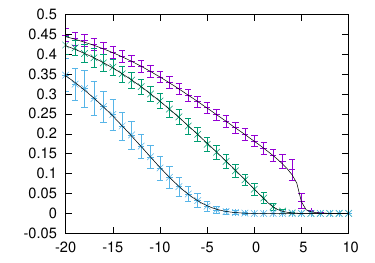}\\
    (c) $\mathrm{BER}_m$\label{fig:sub_2c}
  \end{minipage}%
  \hfill
  \begin{minipage}{0.48\textwidth}
    \centering
    \includegraphics[width=\textwidth]{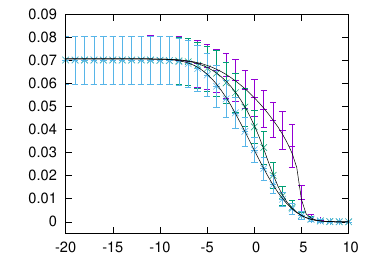}\\
    (d) $\mathrm{BER}_R$\label{fig:sub_2d}
  \end{minipage}

  \caption{(Color online) Order parameters and bit error rates for $\alpha_0 = 1.5$, $\alpha = 1.5$, $N = 1024$, averaged over 200 samples: (a) $m$, (b) $R$, (c) $\mathrm{BER}_m$, and (d) $\mathrm{BER}_R$.
    %Solid black curves indicate theoretical RS predictions, while markers represent MCMC simulation results for $\beta = 1.0$ (purple), $\beta = 0.5$ (green), and $\beta = 0.125$ (light blue).
    Solid curves indicate theoretical RS predictions, while markers represent MCMC simulation results for $\beta = 1.0$
    , $\beta = 0.5$ , and $\beta = 0.125$.
From top to bottom: (a), (b) $\beta = 0.125$, $\beta = 0.5$, and $\beta = 1.0$; (c), (d) $\beta = 1.0$, $\beta = 0.5$, and $\beta = 0.125$.
%    (a), (b) From top to bottom: $\beta = 0.125$,  $\beta = 0.5$, and  $\beta = 1.0$. (c), (d) From top to bottom: $\beta = 1.0$, $\beta = 0.5$, and $\beta = 0.125$.
    Error bars indicate standard deviations.}
  \label{fig.2}
\end{figure}

Next, in Fig.~\ref{fig.2}, we compare the analytical RS predictions with MCMC  simulation results for the order parameters $m$ and $R$, as well as the bit error rates $\mathrm{BER}_m$ and $\mathrm{BER}_R$.
The definitions of the quenched-averaged bit error rates, together with their expressions under the RS ansatz, are given as follows:
\begin{eqnarray}
\mathrm{BER}_m &\equiv& \frac{1}{2}\left(1 - \left[ \frac{1}{K}\sum_{i=1}^K s_i \mathrm{sgn}(\widehat{x}_i)\right]_{\mathrm{av}}\right) = H\left(\frac{\widehat{m}}{\sqrt{\widehat{q}}}\right), \label{eq:ber_m} \\
\mathrm{BER}_R &\equiv& \frac{1}{2}\left(1 - \left[\frac{1}{N} \sum_{\mu=1}^N f_\mu \mathrm{sgn}(\widehat{g}_\mu)\right]_{\mathrm{av}}\right) \nonumber \\
&=& \frac{1}{2}\left\{1 - \tanh \widehat{r}_0 + \frac{1}{\cosh \widehat{r}_0} \left[ e^{\widehat{r}_0} H\left(\frac{\widehat{r}+\widehat{R}}{\sqrt{\widehat{Q}}}\right) - e^{-\widehat{r}_0} H\left(\frac{\widehat{r}-\widehat{R}}{\sqrt{\widehat{Q}}}\right)\right]\right\}, \label{eq:ber_r}
\end{eqnarray}
where $H(x) \equiv \int_x^\infty Dt$ and $[\,\cdot\,]_{\mathrm{av}}$ represents the quenched average defined in Eq.~\eqref{eq:config_av}. 
  Theoretical RS curves and MCMC simulation results show excellent agreement across all tested parameters.

%%%%%%%%%%%%%%%
\subsection{Mismatched hyperparameter case: $\alpha \neq \alpha_0$}

\begin{figure}[tb]
  \centering
  \begin{minipage}{0.48\textwidth}
    \centering
    \includegraphics[width=\textwidth]{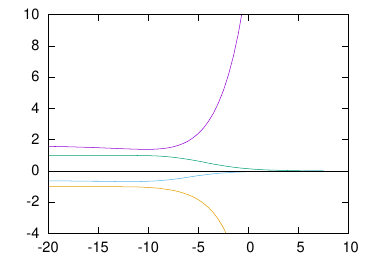}\\
    (a) All $\lambda$ (Type I)\label{fig:sub_3a}
  \end{minipage}%
  \hfill
  \begin{minipage}{0.48\textwidth}
    \centering
    \includegraphics[width=\textwidth]{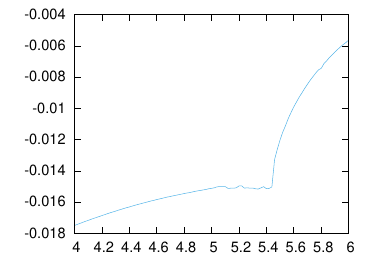}\\
    (b) Enlarged $\lambda_3$ (Type I)\label{fig:sub_3b}
  \end{minipage}

  \vspace{1em}

  \begin{minipage}{0.48\textwidth}
    \centering
    \includegraphics[width=\textwidth]{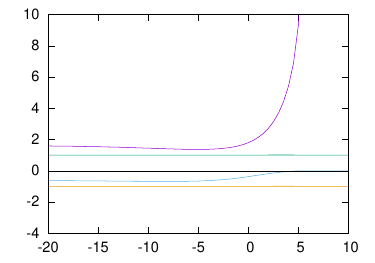}\\
    (c) All $\lambda$ (Type III)\label{fig:sub_3c}
  \end{minipage}%
  \hfill
  \begin{minipage}{0.48\textwidth}
    \centering
    \includegraphics[width=\textwidth]{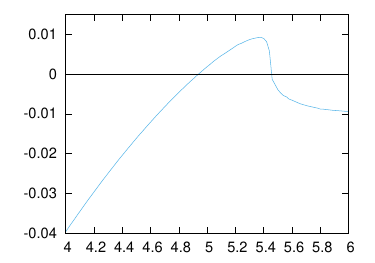}\\
    (d) Enlarged $\lambda_3$ (Type III)\label{fig:sub_3d}
  \end{minipage}
  \caption{(Color online) Eigenvalue behavior for mismatched hyperparameters $\alpha_0 = 1.5$, $\alpha = 3.0$, and $\beta = 1.0$. Color conventions are identical to Fig.~\ref{fig.1}. (a), (b) Eigenvalues for Type I vectors; (c), (d) eigenvalues for Type III vectors. (a), (c) Overview of all eigenvalues; (b), (d) enlarged view of $\lambda_3$.}
  \label{fig.3}
\end{figure}

\begin{figure}[tb]
  \centering
  \begin{minipage}{0.48\textwidth}
    \centering
    \includegraphics[width=\textwidth]{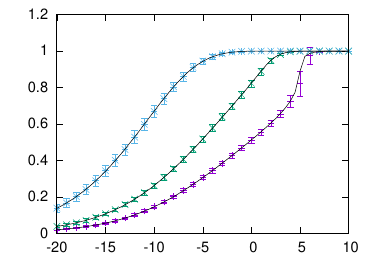}\\
    (a) Order parameter $m$\label{fig:sub_4a}
  \end{minipage}%
  \hfill
  \begin{minipage}{0.48\textwidth}
    \centering
    \includegraphics[width=\textwidth]{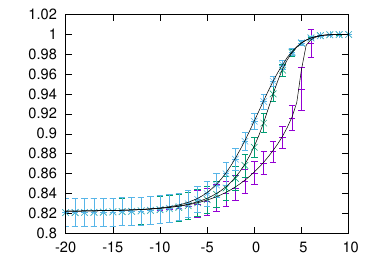}\\
    (b) Order parameter $R$\label{fig:sub_4b}
  \end{minipage}

  \vspace{1em}

  \begin{minipage}{0.48\textwidth}
    \centering
    \includegraphics[width=\textwidth]{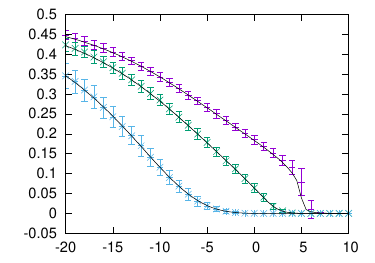}\\
    (c) $\mathrm{BER}_m$\label{fig:sub_4c}
  \end{minipage}%
  \hfill
  \begin{minipage}{0.48\textwidth}
    \centering
    \includegraphics[width=\textwidth]{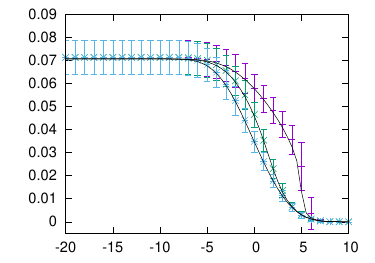}\\
    (d) $\mathrm{BER}_R$\label{fig:sub_4d}
  \end{minipage}

  \caption{(Color online) Order parameters and bit error rates in the AT-stable region for $\alpha_0 = 1.5$, $\alpha = 2.0$, $N = 2048$, averaged over 200 samples: (a) $m$, (b) $R$, (c) $\mathrm{BER}_m$, and (d) $\mathrm{BER}_R$. Curve and marker conventions are identical to Fig.~\ref{fig.2}.}
  \label{fig.4}
\end{figure}

Next, we investigate the effect of hyperparameter mismatch by fixing the true prior parameter $\alpha_0 = 1.5$ while varying the estimator parameter $\alpha \in \{1.0, 1.2, 2.0, 3.0\}$. We evaluated the stability for embedding rates $\beta = 0.125, 0.5$, and $1.0$.
%%%
For moderate mismatch ($\alpha = 1.0, 1.2$, and $2.0$), the RS solution remains AT-stable throughout the evaluated noise regime, and its eigenvalue plots are omitted for brevity. However, for a larger mismatch of $\alpha = 3.0$, the eigenvalue $\lambda_3$ of the Type III mode turns positive in a specific noise region, indicating the  breakdown of the RS stability. Figure~\ref{fig.3} illustrates the eigenvalue behavior for $\alpha_0 = 1.5, \alpha = 3.0$, and $\beta = 1.0$. As seen in Fig.~\ref{fig.3}~(d), $\lambda_3$ crosses zero and becomes positive near $E_b/N_0 = 5.2\ \mathrm{dB}$.
We then compare the MCMC simulation results with the theoretical RS predictions under hyperparameter mismatch. In the AT-stable regime (e.g., $\alpha = 2.0$), the analytical predictions match the simulation results remarkably well for both order parameters and bit error rates across all tested $\beta$ values, as shown in Fig.~\ref{fig.4}.
%%%
Conversely, when the system enters the AT-unstable regime ($\alpha = 3.0$), directly identifying these deviations from overview plots such as Fig. 5 is difficult due to the scale.
To inspect the discrepancy clearly, we highlight the fine-scale behavior of the bit error rates across the  AT-unstable region for a large system size ($N=10000$).
Figure~\ref{fig.6} presents the AT-stable case ($\alpha = 2.0$), showing excellent agreement between theory and simulation. In contrast, Fig.~\ref{fig.7} illustrates the AT-unstable case ($\alpha = 3.0$), where systematic discrepancies between the RS predictions and MCMC simulation results emerge in the vicinity of the AT-unstable region.

\begin{figure}[tb]
  \centering
  \begin{minipage}{0.48\textwidth}
    \centering
    \includegraphics[width=\textwidth]{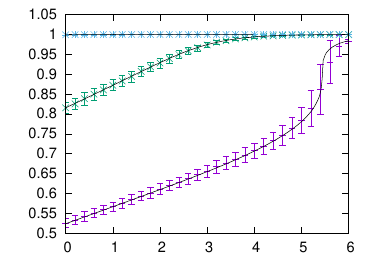}\\
    (a) Order parameter $m$\label{fig:sub_5a}
  \end{minipage}%
  \hfill
  \begin{minipage}{0.48\textwidth}
    \centering
    \includegraphics[width=\textwidth]{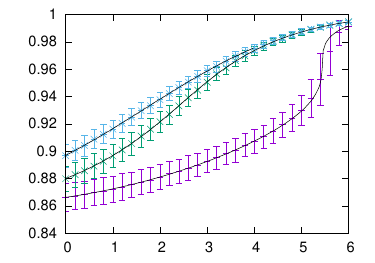}\\
    (b) Order parameter $R$\label{fig:sub_5b}
  \end{minipage}

  \vspace{1em}

  \begin{minipage}{0.48\textwidth}
    \centering
    \includegraphics[width=\textwidth]{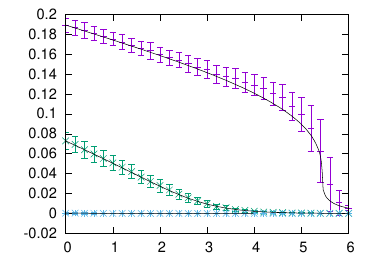}\\
    (c) $\mathrm{BER}_m$\label{fig:sub_5c}
  \end{minipage}%
  \hfill
  \begin{minipage}{0.48\textwidth}
    \centering
    \includegraphics[width=\textwidth]{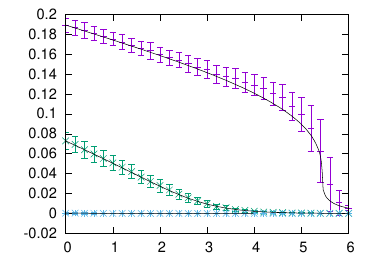}\\
    (d) $\mathrm{BER}_R$\label{fig:sub_5d}
  \end{minipage}

  \caption{(Color online) Order parameters and bit error rates for $\alpha_0 = 1.5$, $\alpha = 3.0$, $N = 4096$, averaged over 200 samples: (a) $m$, (b) $R$, (c) $\mathrm{BER}_m$, and (d) $\mathrm{BER}_R$. Curve and marker conventions are identical to Fig.~\ref{fig.2}.}
  \label{fig.5}
\end{figure}

\begin{figure}[tb]
  \centering
  \begin{minipage}{0.48\textwidth}
    \centering
    \includegraphics[width=\textwidth]{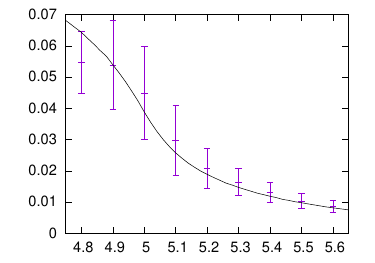}\\
    (a) $\mathrm{BER}_m$\label{fig:sub_6a}
  \end{minipage}%
  \hfill
  \begin{minipage}{0.48\textwidth}
    \centering
    \includegraphics[width=\textwidth]{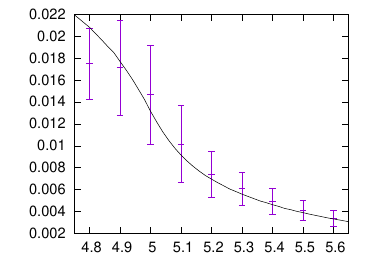}\\
    (b) $\mathrm{BER}_R$\label{fig:sub_6b}
  \end{minipage}

  \caption{(Color online) Detailed bit error rate behavior in the AT-stable region for $\alpha_0 = 1.5$, $\alpha = 2.0$, $\beta = 1.0$, $N = 10000$, averaged over 200 samples: (a) $\mathrm{BER}_m$ and (b) $\mathrm{BER}_R$.}
  \label{fig.6}
\end{figure}

\begin{figure}[tb]
  \centering
  \begin{minipage}{0.48\textwidth}
    \centering
    \includegraphics[width=\textwidth]{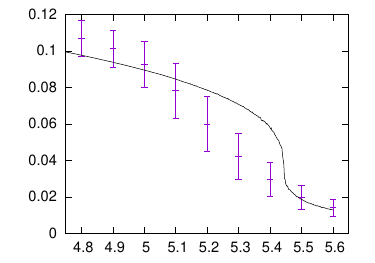}\\
    (a) $\mathrm{BER}_m$\label{fig:sub_7a}
  \end{minipage}%
  \hfill
  \begin{minipage}{0.48\textwidth}
    \centering
    \includegraphics[width=\textwidth]{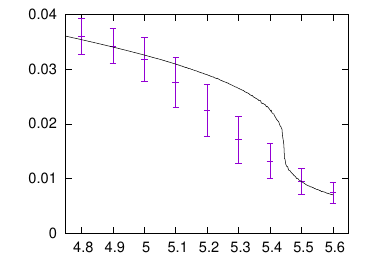}\\
    (b) $\mathrm{BER}_R$\label{fig:sub_7b}
  \end{minipage}

  \caption{(Color online) Detailed bit error rate behavior near the AT-unstable region for $\alpha_0 = 1.5$, $\alpha = 3.0$, $\beta = 1.0$, $N = 10000$, averaged over 200 samples: (a) $\mathrm{BER}_m$ and (b) $\mathrm{BER}_R$.}
  \label{fig.7}
\end{figure}

\section{CONCLUSION}
\label{sec:conclusion}
In this paper, based on a self-contained formulation of the hybrid watermarking
and image restoration model, we investigated the AT stability of the RS solution
in a hybrid system that simultaneously performs message decoding and image reconstruction.
Extending our previous Bayes-estimation-based watermarking framework, we analyzed the stability of the RS solution
 against RSB by evaluating the replicon Hessian matrix ${\cal H}_{\mathrm{rp}}$.
%%%
 Specifically, we derived the characteristic quartic algebraic equation governing  the eigenvalues of ${\cal H}_{\mathrm{rp}}$ for the RS solution.
 
By classifying the replicon modes into Type I, Type II, and Type III eigenvectors,
we demonstrated that the AT stability criterion remains satisfied for Type I and Type II modes
across all evaluated parameter regimes.
In contrast, the Type III mode undergoes the stability breakdown in specific parameter regions
when hyperparameter mismatch is introduced.
During this analysis, we clarified the structure of the total replicon eigenspace as
\begin{equation}
\mathcal{V}_{\mathrm{rp}}^{\mathrm{total}} = \mathbb{R}^4 \otimes \mathcal{V}_{\mathrm{rp}},
\end{equation}
where $\mathcal{V}_{\mathrm{rp}}^{\mathrm{total}}$, $\mathbb{R}^4$, and $\mathcal{V}_{\mathrm{rp}}$ 
denote the total replicon eigenspace, the 4-dimensional space associated with the matrix $C$, 
and the eigenspace spanned by the Type I, II, and III vectors, respectively. 
This yields a total dimension of $4n_1$, as expected.

By solving the characteristic quartic equation at the SPE solution
 and comparing  the theoretical results with MCMC simulations, we confirmed that the parameter regime
where the MCMC results deviate from the theoretical RS predictions largely coincides with the AT-unstable region.
This finding indicates that RSB must be explicitly incorporated into
the theoretical analysis in order to accurately describe the system's behavior
when the AT stability condition is violated.

Future work includes deriving the one-step replica symmetry breaking (1RSB) solution \cite{MPV1987}
to quantitatively predict the system's performance within the AT-unstable region, as well as extending numerical calculations and MCMC simulations to channel noise mismatches. 

\begin{acknowledgment}

%  T. U. expresses deep gratitude to Professor So Kitsunezaki of Nara Women's University for his invaluable support,
%  particularly for the opportunity to use the workstations in his laboratory
%  to perform numerical calculations and computationally intensive simulations.

  T. U. expresses deep gratitude to Professor So Kitsunezaki of Nara Women's University for his invaluable support,
  particularly for the opportunity to use the workstations in his laboratory
  to conduct numerical calculations and computationally intensive simulations.
\end{acknowledgment}  

\appendix
\section{Derivation of free energy}
\label{sec:app_free_energy}

In this appendix, we derive the quenched average of the $n$-th power of the partition function, $[Z^n]_{\mathrm{av}}$.
By introducing the conjugate variable $\widehat{q}_{ab}$ corresponding to the order parameter $q_{ab}$, we obtain the identity:
\begin{equation}
  1 = \int \frac{iK \mathrm{d}q_{ab} \mathrm{d}\widehat{q}_{ab}}{2\pi} \exp\left[-K\widehat{q}_{ab} \left(q_{ab} - \frac{1}{K}\sum_{i=1}^K x_i^a x_i^b \right) \right].
\label{eq:delta-function}  
\end{equation}
Analogous identities are introduced for $m_a$, $v_0^\mu$, and $v_a^\mu$ with their respective conjugate variables.
Using these representations, $[Z^n]_{\mathrm{av}}$ defined in Eq.~\eqref{eq:npowerofZ} is expressed as
\begin{align}
[Z^n]_{\mathrm{av}} &= \frac{1}{2^K Z_f} \int \prod_{a<b} \left[\frac{iK \mathrm{d}\widehat{q}_{ab}\mathrm{d}q_{ab}}{2\pi}\right] \int \prod_a \left[\frac{iK \mathrm{d}\widehat{m}_a \mathrm{d}m_a}{2\pi}\right] \int \prod_\mu \left[\frac{\mathrm{d}\widehat{v}_0^\mu \mathrm{d}v_0^\mu}{2\pi}\right] \int \prod_{a,\mu} \left[\frac{\mathrm{d}\widehat{v}_a^\mu \mathrm{d}v_a^\mu}{2\pi}\right] \nonumber \\
&\quad \times \exp\left[-K\sum_{a<b}\widehat{q}_{ab} q_{ab} - K\sum_a \widehat{m}_a m_a + i\sum_\mu \widehat{v}_0^\mu v_0^\mu + i\sum_{a,\mu}\widehat{v}_a^\mu v_a^\mu\right] \nonumber \\
&\quad \times (2\pi\sigma_0^2)^{-\frac{N}{2}} \int \left[\prod_{\mu=1}^N \mathrm{d}\widetilde{f}_\mu\right] \mathrm{Tr}_{\vec{s}} \mathrm{Tr}_{\{\vec{x}^a\}} \mathrm{Tr}_{\vec{f}} \mathrm{Tr}_{\{\vec{g}^a\}} \nonumber \\
&\quad \times \left\langle \exp\left[-\frac{1}{2\sigma_0^2}\sum_{\mu=1}^N (\widetilde{f}_\mu - v_0^\mu - f_\mu)^2 - i\sum_\mu \widehat{v}_0^\mu \frac{1}{\sqrt{K}}\sum_{i=1}^K \xi_i^\mu s_i - i\sum_{a,\mu}\widehat{v}_a^\mu \frac{1}{\sqrt{K}}\sum_{i=1}^K \xi_i^\mu x_i^a \right.\right. \nonumber \\
&\quad \left.\left. + \sum_{a<b}\widehat{q}_{ab}\sum_{i=1}^K x_i^a x_i^b + \sum_a \widehat{m}_a\sum_{i=1}^K s_i x_i^a - \frac{1}{2\sigma^2}\sum_a \sum_{\mu=1}^N (\widetilde{f}_\mu - v_a^\mu - g_\mu^a)^2 \right] \right\rangle_\xi \psi(\vec{f}, \{\vec{g}^a\}),
\label{eq:Zpowern-1}
\end{align}
where the  factor $\psi(\vec{f}, \{\vec{g}^a\})$ is given by
\begin{equation}
\psi(\vec{f}, \{\vec{g}^a\}) = \frac{\alpha_0}{2N}\left(\sum_{\mu=1}^N f_\mu\right)^2 + \frac{\alpha}{2N}\sum_a \left(\sum_{\mu=1}^N g_\mu^a\right)^2.
\end{equation}
%%%
Averaging over the quenched variables $\xi_i^\mu$  for $K \gg 1$ yields 
\begin{align}
&\left\langle \exp\left[ -i\frac{1}{\sqrt{K}}\sum_{i=1}^K \sum_{\mu=1}^N \left(\widehat{v}_0^\mu s_i + \sum_{a=1}^n \widehat{v}_a^\mu x_i^a \right)\xi_i^\mu \right] \right\rangle_\xi \nonumber \\
&\qquad \simeq \exp\left[-\sum_{\mu=1}^N \left(\frac{1}{2}(\widehat{v}_0^\mu)^2 + \frac{1}{2}\sum_{a=1}^n (\widehat{v}_a^\mu)^2 + \sum_{a<b} q_{ab}\widehat{v}_a^\mu \widehat{v}_b^\mu + \widehat{v}_0^\mu \sum_{a=1}^n m_a \widehat{v}_a^\mu \right)\right].
\end{align}
%%%
We decompose the integrand of $[Z^n]_{\mathrm{av}}$ into three distinct contributions: $e^{NG_1}$ collects terms involving only the message order parameters;
$e^{NG_2}$ contains the binary message variables $s$ and $\{x_a\}$; while $e^{NG_3}$  consists of all remaining variables.  
Consequently, Eq.~\eqref{eq:Zpowern-1} simplifies to
\begin{align}
[Z^n]_{\mathrm{av}} &= \frac{1}{2^K Z_f} \int \prod_{a<b} \left[\frac{iK \mathrm{d}\widehat{q}_{ab}\mathrm{d}q_{ab}}{2\pi}\right] \int \prod_a \left[\frac{iK \mathrm{d}\widehat{m}_a \mathrm{d}m_a}{2\pi}\right] \int \prod_\mu \left[\frac{\mathrm{d}\widehat{v}_0^\mu \mathrm{d}v_0^\mu}{2\pi}\right] \int \prod_{a,\mu} \left[\frac{\mathrm{d}\widehat{v}_a^\mu \mathrm{d}v_a^\mu}{2\pi}\right] \nonumber \\
&\quad \times (2\pi\sigma_0^2)^{-\frac{N}{2}} \int \left[\prod_{\mu=1}^N \mathrm{d}\widetilde{f}_\mu\right] \mathrm{Tr}_{\vec{s}} \mathrm{Tr}_{\{\vec{x}^a\}} \mathrm{Tr}_{\vec{f}} \mathrm{Tr}_{\{\vec{g}^a\}} \exp[N(G_1 + G_2 + G_3)].
\end{align}
Here, each component is evaluated as follows. First, for $G_1$ and $G_2$, we have
\begin{align}
G_1 &= -\beta \sum_{a<b}\widehat{q}_{ab} q_{ab} - \beta \sum_a \widehat{m}_a m_a, \\
e^{NG_2} &= \mathrm{Tr}_{\vec{s}} \mathrm{Tr}_{\{\vec{x}^a\}} \exp\left[\sum_{a<b}\widehat{q}_{ab}\sum_{i=1}^K x_i^a x_i^b + \sum_a \widehat{m}_a\sum_{i=1}^K s_i x_i^a\right], \label{eq:expg2}
\end{align}
where $\beta \equiv K/N$ is the embedding rate. Since the right-hand side of Eq.~\eqref{eq:expg2} factorizes site-wise with respect to index $i$, it simplifies to
\begin{align}
e^{NG_2} &= \left(\mathrm{Tr}_s \mathrm{Tr}_{\{x^a\}} \exp\left[\sum_{a<b}\widehat{q}_{ab} x^a x^b + \sum_a \widehat{m}_a s x^a\right]\right)^K, \\
G_2 &= \beta \ln \left(\mathrm{Tr}_s \mathrm{Tr}_{\{x^a\}} \exp\left[\sum_{a<b}\widehat{q}_{ab} x^a x^b + \sum_a \widehat{m}_a s x^a\right]\right),
\end{align}
where $\mathrm{Tr}_s$ and $\mathrm{Tr}_{\{x^a\}}$ denote the summation over $s \in \{-1, +1\}$ and that over $x^a \in \{-1, +1\}$ for $a = 1, \dots, n$, respectively.
Second, for $G_3$, the corresponding exponential term $e^{NG_3}$ is written as
\begin{align}
e^{NG_3} &= \mathrm{Tr}_{\vec{f}} \mathrm{Tr}_{\{\vec{g}^a\}} \prod_\mu \left\{ \int \frac{\mathrm{d}\widehat{v}_0^\mu \mathrm{d}v_0^\mu}{2\pi} \int \prod_a \left[\frac{\mathrm{d}\widehat{v}_a^\mu \mathrm{d}v_a^\mu}{2\pi}\right] \int \frac{\mathrm{d}\widetilde{f}_\mu}{\sqrt{2\pi}\sigma_0} \right. \nonumber \\
&\quad \times \exp\left[ i\widehat{v}_0^\mu v_0^\mu + i\sum_a \widehat{v}_a^\mu v_a^\mu - \frac{1}{2}(\widehat{v}_0^\mu)^2 - \frac{1}{2}\sum_{a=1}^n (\widehat{v}_a^\mu)^2 - \sum_{a<b} q_{ab}\widehat{v}_a^\mu \widehat{v}_b^\mu - \widehat{v}_0^\mu \sum_{a=1}^n m_a \widehat{v}_a^\mu \right. \nonumber \\
&\quad \left.\left. - \frac{1}{2\sigma_0^2}(\widetilde{f}_\mu - v_0^\mu - f_\mu)^2 - \frac{1}{2\sigma^2}\sum_a (\widetilde{f}_\mu - v_a^\mu - g_\mu^a)^2 \right] \right\} \exp[\psi(\vec{f}, \{\vec{g}^a\})].
\end{align}
%%%
The image-related order parameters $R_a, r_0, r_a$, and $Q_{ab}$ are introduced via delta functions. By inserting the integral identity for $R_a$,
\begin{equation}
1 = \int \frac{iN \mathrm{d}R_a \mathrm{d}\widehat{R}_a}{2\pi} \exp\left[-\widehat{R}_a\left(NR_a - \sum_{\mu=1}^N f_\mu g_\mu^a\right)\right],
\end{equation}
and similar representations for $r_0, r_a$, and $Q_{ab}$, the factor $\exp[\psi(\vec{f}, \{\vec{g}^a\})]$ is
replaced by $e^{NG_4}$, 
\begin{align}
  %\exp[\psi(\vec{f}, \{\vec{g}^a\})] &= \int \mathrm{d}\vec{X}
  \exp[NG_4]& \equiv \exp\left[-N\left(\widehat{r}_0 r_0 + \sum_a \widehat{r}_a r_a + \sum_a \widehat{R}_a R_a +
    \sum_{a<b}\widehat{Q}_{ab} Q_{ab}\right)\right] \nonumber \\
&\quad \times \left[\prod_\mu \exp\left(\widehat{r}_0 f_\mu + \sum_a \widehat{r}_a g_\mu^a + \sum_a \widehat{R}_a f_\mu g_\mu^a + \sum_{a<b}\widehat{Q}_{ab} g_\mu^a g_\mu^b\right)\right] \nonumber \\
&\quad \times \exp\left[\frac{N}{2}\left(\alpha_0 r_0^2 + \alpha \sum_a r_a^2\right)\right].
\end{align}
By defining  the integration measure $\mathrm{d}\vec{X}$ as
\begin{equation}
\mathrm{d}\vec{X} \equiv \left[\prod_a \int \frac{iN \mathrm{d}R_a \mathrm{d}\widehat{R}_a}{2\pi}\right] \int \frac{iN \mathrm{d}r_0 \mathrm{d}\widehat{r}_0}{2\pi} \left[\prod_a \int \frac{iN \mathrm{d}r_a \mathrm{d}\widehat{r}_a}{2\pi}\right] \left[\prod_{a<b} \int \frac{iN \mathrm{d}Q_{ab}\mathrm{d}\widehat{Q}_{ab}}{2\pi}\right],
\end{equation}
%%%
 $e^{NG_3}$ becomes
\begin{align}
e^{NG_3} &= \int \mathrm{d}\vec{X} \prod_\mu \left\{ \mathrm{Tr}_{f_\mu} \mathrm{Tr}_{\{g_\mu^a\}} \int \frac{\mathrm{d}\widehat{v}_0^\mu \mathrm{d}v_0^\mu}{2\pi} \int \prod_a \left[\frac{\mathrm{d}\widehat{v}_a^\mu \mathrm{d}v_a^\mu}{2\pi}\right] \int \frac{\mathrm{d}\widetilde{f}_\mu}{\sqrt{2\pi}\sigma_0} \right. \nonumber \\
&\quad \times \exp\left[ i\widehat{v}_0^\mu v_0^\mu + i\sum_a \widehat{v}_a^\mu v_a^\mu - \frac{1}{2}(\widehat{v}_0^\mu)^2 - \frac{1}{2}\sum_{a=1}^n (\widehat{v}_a^\mu)^2
  %+ \widehat{r}_0  f_\mu + \sum_a \widehat{r}_a g_\mu^a \right. \nonumber \\
  %\quad + \sum_a \widehat{R}_a f_\mu g_\mu^a + \sum_{a<b}\widehat{Q}_{ab} g_\mu^a g_\mu^b
  - \sum_{a<b} q_{ab}\widehat{v}_a^\mu \widehat{v}_b^\mu - \widehat{v}_0^\mu \sum_{a=1}^n m_a \widehat{v}_a^\mu
  \right. \nonumber \\
&\quad \left.\left. - \frac{1}{2\sigma_0^2}(\widetilde{f}_\mu - v_0^\mu - f_\mu)^2 - \frac{1}{2\sigma^2}\sum_a (\widetilde{f}_\mu - v_a^\mu - g_\mu^a)^2 \right] \right\} \exp \left[ N G_4 \right].
%&\quad \times \exp\left[-N\left(\widehat{r}_0 r_0 + \sum_a \widehat{r}_a r_a + \sum_a \widehat{R}_a R_a + \sum_{a<b}\widehat{Q}_{ab} Q_{ab}\right) + \frac{N}{2}\left(\alpha_0 r_0^2 + \alpha \sum_a r_a^2\right)\right].
\end{align}
Since the integrand factorizes over index $\mu$, we define $\widetilde{G}_3$ such that
\begin{align}
e^{NG_3} &= \int \mathrm{d}\vec{X} \, e^{N\widetilde{G}_3}, \\
e^{\widetilde{G}_3} &\equiv \mathrm{Tr}_f \mathrm{Tr}_{\{g^a\}} \int \frac{\mathrm{d}\widehat{v}_0 \mathrm{d}v_0}{2\pi} \int \prod_a \left[\frac{\mathrm{d}\widehat{v}_a \mathrm{d}v_a}{2\pi}\right] \int \frac{\mathrm{d}\widetilde{f}}{\sqrt{2\pi}\sigma_0} \nonumber \\
&\quad \times \exp\left[ i\widehat{v}_0 v_0 + i\sum_a \widehat{v}_a v_a - \frac{1}{2}(\widehat{v}_0)^2 - \frac{1}{2}\sum_{a=1}^n (\widehat{v}_a)^2 - \sum_{a<b} q_{ab}\widehat{v}_a \widehat{v}_b - \widehat{v}_0 \sum_{a=1}^n m_a \widehat{v}_a \right. \nonumber \\
&\quad \left. - \frac{1}{2\sigma_0^2}(\widetilde{f} - v_0 - f)^2 - \frac{1}{2\sigma^2}\sum_a (\widetilde{f} - v_a - g^a)^2 + G_4 \right], \\
G_4 &= -\left(\widehat{r}_0 r_0 + \sum_a \widehat{r}_a r_a + \sum_a \widehat{R}_a R_a + \sum_{a<b}\widehat{Q}_{ab} Q_{ab}\right) + \frac{1}{2}\left(\alpha_0 r_0^2 + \alpha \sum_a r_a^2\right) \nonumber \\
&\quad + \widehat{r}_0 f + \sum_a \widehat{r}_a g^a + \sum_a \widehat{R}_a f g^a + \sum_{a<b}\widehat{Q}_{ab} g^a g^b. \label{eq:g4}
\end{align}
%%%
Performing Gaussian integrations in $e^{\widetilde{G}_3}$ with respect to $\widetilde{f}, v_0$, and $\widehat{v}_0$ yields
\begin{align}
e^{\widetilde{G}_3} &= \mathrm{Tr}_f \mathrm{Tr}_{\{g^a\}} \sqrt{\frac{n}{\sigma^2 + n(1 + \sigma_0^2)}} \exp\left[ \frac{n}{2(\sigma^2 + n(1 + \sigma_0^2))} \left(\frac{i}{n}\bar{B} - \sum_a m_a \widehat{v}_a - if\right)^2 
\right. \nonumber \\ &\quad \left.
+ \frac{\bar{B}^2}{2n(\sigma^2 + n\sigma_0^2)} \right] %\nonumber \\ &\quad \times
\left[\prod_a \int \frac{\mathrm{d}v_a \mathrm{d}\widehat{v}_a}{2\pi}\right] \frac{\sigma}{\sqrt{\sigma^2 + n\sigma_0^2}} \exp\left[ -\frac{1}{2\sigma^2}\sum_a (b^a)^2 \right] \nonumber \\
&\quad \times \sqrt{\frac{\sigma^2 + n\sigma_0^2}{n}} \exp\left[ \frac{\sigma_0^2}{2(\sigma^2 + n\sigma_0^2)\sigma^2}\bar{B}^2 + i\sum_a \widehat{v}_a v_a - \frac{1}{2}\sum_a (\widehat{v}_a)^2 - \sum_{a<b} q_{ab}\widehat{v}_a \widehat{v}_b + G_4 \right],
\end{align}
where $\bar{B} \equiv \sum_a b^a$ with $b^a \equiv v_a + g^a$.
We define $\xi \equiv \frac{1}{\sigma}\sqrt{\frac{1 + \sigma_0^2}{\sigma^2 + n(1 + \sigma_0^2)}}$, whose square represents the coefficient of $\bar{B}^2 / 2$.
Applying the following Hubbard-Stratonovich transformation:
\begin{equation}
\exp\left[\frac{1}{2}\xi^2 \bar{B}^2\right] = \int Dt \exp[\xi \bar{B} t],
\end{equation}
where $Dt \equiv \frac{\mathrm{d}t}{\sqrt{2\pi}} e^{-t^2/2}$, we obtain
\begin{align}
  e^{\widetilde{G}_3} &= \mathrm{Tr}_f \mathrm{Tr}_{\{g^a\}} \frac{\sigma}{\sqrt{\sigma^2 + n(1 + \sigma_0^2)}} \int Dt
  \int \left[\prod_a  \frac{\mathrm{d}v_a \mathrm{d}\widehat{v}_a}{2\pi}\right] \nonumber \\
&\quad \times \exp\left[ \xi t \bar{B} - \frac{i}{\sigma^2 + n(1 + \sigma_0^2)}\bar{B}\left(\sum_a m_a \widehat{v}_a + if\right) 
\right. \nonumber \\ &\quad \left.
+ \frac{n}{2(\sigma^2 + n(1+\sigma_0^2))} \left(\sum_a m_a \widehat{v}_a + if\right)^2 \right. \nonumber \\
&\quad \left. - \frac{1}{2\sigma^2}\sum_a (b^a)^2 + i\sum_a \widehat{v}_a v_a - \frac{1}{2}\sum_a (\widehat{v}_a)^2 - \sum_{a<b} q_{ab}\widehat{v}_a \widehat{v}_b + G_4 \right].
\end{align}
%%%
Integrating over $b^a$ instead of $v_a$, we arrive at
\begin{align}
e^{\widetilde{G}_3} &= \mathrm{Tr}_f \mathrm{Tr}_{\{g^a\}} \frac{\sigma^{n+1}}{\sqrt{\sigma^2 + n(1 + \sigma_0^2)}} \int Dt \left[\prod_a \int \frac{\mathrm{d}\widehat{v}_a}{\sqrt{2\pi}}\right] e^{{\cal L} + G_4}, \label{eq:g3-1} \\
{\cal L} &= \frac{\sigma^2}{2} n(\xi t)^2 - \frac{\sigma^2}{2}\sum_a (\widehat{v}_a)^2 - in\xi t U \frac{\sigma^2}{\sigma^2 + n(1 + \sigma_0^2)} + i\sigma^2 \xi t \sum_a \widehat{v}_a \nonumber \\
&\quad + U \frac{\sigma^2}{\sigma^2 + n(1 + \sigma_0^2)}\sum_a \widehat{v}_a - i\sum_a \widehat{v}_a g^a - \frac{1}{2}\sum_a (\widehat{v}_a)^2 - \sum_{a<b} q_{ab}\widehat{v}_a \widehat{v}_b + {\cal O}(n^2), \label{eq:g3-2}
\end{align}
where $U \equiv \sum_a m_a \widehat{v}_a + if$.
Consequently, $[Z^n]_{\mathrm{av}}$ is expressed as
\begin{align}
[Z^n]_{\mathrm{av}} &= \int \mathrm{d}\vec{V} e^{NG}, \\
G &= G_1 + G_2 + \widetilde{G}_3 - \beta \ln 2 - \frac{1}{N}\ln Z_f, \\
\mathrm{d}\vec{V} &\equiv \left[\prod_{a<b} \frac{iK \mathrm{d}\widehat{q}_{ab}\mathrm{d}q_{ab}}{2\pi}\right] \left[\prod_a \frac{iK \mathrm{d}\widehat{m}_a \mathrm{d}m_a}{2\pi}\right] \mathrm{d}\vec{X}.
\end{align}
Equations~\eqref{eq:g4}, \eqref{eq:g3-1}, and \eqref{eq:g3-2} serve as the starting point for calculating
the Hessian matrix of $G_3$ to investigate the AT stability.

%%%%%%%%%%%%%%%%%%%%%%%%%%%%%%%%%%%%%%%%
\section{RS solution}
\label{sec:RS_solution}
Under the RS ansatz, we assume that the order parameters are independent of the replica indices, i.e.,
\begin{align}
m_a &= m, \quad q_{ab} = q, \quad \widehat{m}_a = \widehat{m}, \quad \widehat{q}_{ab} = \widehat{q}, \\
R_a &= R, \quad Q_{ab} = Q, \quad \widehat{R}_a = \widehat{R}, \quad \widehat{Q}_{ab} = \widehat{Q}, \quad r_a = r, \quad \widehat{r}_a = \widehat{r}.
\end{align}
By neglecting terms of order $n^2$ and higher, the function $G$ is decomposed into terms of order $n^0$ and $n^1$, expressed as $G = G^0 + nG^1$. Here, $G$ represents $G_{1,\mathrm{RS}}$, $G_{2,\mathrm{RS}}$, and $G_{3,\mathrm{RS}}$
 for the RS solution:
\begin{align}
G_{\mathrm{RS}} &= G_{1,\mathrm{RS}} + G_{2,\mathrm{RS}} + \widetilde{G}_{3,\mathrm{RS}} - \beta \ln 2 - \frac{1}{N}\ln Z_f, \\
G_{\mathrm{RS}}^0 &= G_{1,\mathrm{RS}}^0 + G_{2,\mathrm{RS}}^0 + \widetilde{G}_{3,\mathrm{RS}}^0 - \beta \ln 2 - \frac{1}{N}\ln Z_f, \\
G_{\mathrm{RS}}^1 &= G_{1,\mathrm{RS}}^1 + G_{2,\mathrm{RS}}^1 + \widetilde{G}_{3,\mathrm{RS}}^1.
\end{align}
%%%
The components $G_{1,\mathrm{RS}}^0$, $G_{1,\mathrm{RS}}^1$, $G_{2,\mathrm{RS}}^0$, and $G_{2,\mathrm{RS}}^1$ are evaluated as
\begin{align}
G_{1,\mathrm{RS}}^0 &= 0, \quad G_{1,\mathrm{RS}}^1 = \frac{1}{2} \beta q \widehat{q} - \beta m \widehat{m}, \\
G_{2,\mathrm{RS}}^0 &= \beta \ln 2, \quad G_{2,\mathrm{RS}}^1 = -\frac{1}{2} \beta \widehat{q} + \beta \int D z \ln \left[ 2 \cosh \left( \sqrt{\widehat{q}} z + \widehat{m} \right) \right].
\end{align}
Next, we evaluate $G_{3,\mathrm{RS}}$. Under the RS ansatz, we denote $\widetilde{G}_3$, $G_4$, and ${\cal L}$ as $\widetilde{G}_{3,\mathrm{RS}}$, $G_{4,\mathrm{RS}}$, and ${\cal L}_{\mathrm{RS}}$, respectively.
First, we perform the integration with respect to $\widehat{v}_a$, which appears only in ${\cal L}_{\mathrm{RS}}$. We rewrite $\sum_{a<b} \widehat{v}_a \widehat{v}_b$ as
\begin{equation}
\sum_{a<b} \widehat{v}_a \widehat{v}_b = \frac{1}{2} V^2 - \frac{1}{2} \sum_a (\widehat{v}_a)^2,
\end{equation}
where $V \equiv \sum_a \widehat{v}_a$. Defining the coefficient of $V^2/2$ in ${\cal L}_{\mathrm{RS}}$ as $\eta^2$, we apply the Hubbard-Stratonovich (H-S) transformation to obtain
\begin{equation}
\exp\left[\frac{1}{2}\eta^2 V^2\right] = \int Dz \exp[\eta V z],
\end{equation}
where $\eta \equiv \sqrt{\frac{2m\sigma^2}{\sigma^2 + n(1+\sigma_0^2)} - q}$. Up to order  $\mathcal{O}(n)$, $\exp[{\cal L}_{\mathrm{RS}}]$ becomes
\begin{align}
\exp[{\cal L}_{\mathrm{RS}}] &= \int Dz \exp\left[ -\frac{\widehat{a}}{2}\sum_a (\widehat{v}_a)^2 + \sum_a \widehat{b}(u, f, g^a) \widehat{v}_a + \frac{\sigma^2}{2} n (\xi t)^2 + n \xi t f  \right], \label{eq:calL} \\
\widehat{a} &\equiv \sigma^2 + 1 - q, \quad u \equiv \eta z + Bt, \quad \widehat{b}(u, f, g^a) \equiv u + i\frac{\sigma^2}{\sigma^2 + n(1+\sigma_0^2)} f - i g^a, \nonumber \\
B &\equiv i\xi (\sigma^2 - n m ). \nonumber
\end{align}
%%%
Integrating over $\widehat{v}_a$, we obtain
\begin{align}
e^{\widetilde{G}_{3,\mathrm{RS}}} &= \mathrm{Tr}_f \mathrm{Tr}_{\{g^a\}} \frac{\sigma^{n+1}}{\sqrt{\sigma^2 + n(1+\sigma_0^2)}} \widehat{a}^{-n/2} \int Dt \int Dz \nonumber \\
&\quad \times \exp\left[ \frac{1}{2\widehat{a}}\sum_a \widehat{b}(u, f, g^a)^2 + \frac{\sigma^2}{2} n (\xi t)^2
  +  \xi t f n + G_{4,\mathrm{RS}} \right].
\end{align}
We then perform the integration with respect to $t$ and $u$.
For any generic integrand $\varphi(t,u)$, we utilize the following Gaussian integration formula:
\begin{quote}
\textbf{Formula:}
\begin{align}
I &\equiv \int_{-\infty}^{\infty} \frac{\mathrm{d}t}{\sqrt{2\pi}} \int_{-\infty}^{\infty} \frac{\mathrm{d}u}{\sqrt{2\pi}} \exp\left[ -\frac{a_1}{2}t^2 + a_2 t + a_3 t u - \frac{a_4}{2}u^2 + a_5 u + a_6 \right] \varphi(t, u) \nonumber \\
&= \frac{1}{\sqrt{a_1 a_4 - a_3^2}} \int Dx_1 \int Dx_2 \, \varphi\left(b_1 x_1 + b_2(b_4 x_2 + b_5) + b_3, \, b_4 x_2 + b_5\right) \nonumber \\
&\quad \times \exp\left[ \frac{(a_1 a_5 + a_2 a_3)^2}{2 a_1(a_1 a_4 - a_3^2)} + \frac{a_2^2}{2 a_1} + a_6 \right], \label{eq:formula}
\end{align}
where $t = b_1 x_1 + b_2(b_4 x_2 + b_5) + b_3$, $u = b_4 x_2 + b_5$, and the coefficients are given by
\begin{equation}
b_1 = \frac{1}{\sqrt{a_1}}, \quad b_2 = \frac{a_3}{a_1}, \quad b_3 = \frac{a_2}{a_1}, \quad b_4 = \sqrt{\frac{a_1}{a_1 a_4 - a_3^2}}, \quad b_5 = \frac{a_1 a_5 + a_2 a_3}{a_1 a_4 - a_3^2}.
\end{equation}
\end{quote}
Matching our integral to $I$, we write
\begin{align}
&\int Dt \int Dz \exp\left[ \frac{1}{2\widehat{a}}\sum_a \widehat{b}(\eta z + Bt, f, g^a)^2 + \frac{\sigma^2}{2} n (\xi t)^2 +  \xi t f n \right] \nonumber \\
  &\quad = \int \frac{\mathrm{d}t}{\sqrt{2\pi}} \int \frac{\mathrm{d}u}{\eta\sqrt{2\pi}} \exp\left[ -\frac{1}{2}t^2 - \frac{1}{2}\left(\frac{u - Bt}{\eta}\right)^2 + \frac{1}{2\widehat{a}}\sum_a \widehat{b}(u, f, g^a)^2 + \frac{\sigma^2}{2} n (\xi t)^2      +  \xi t f n \right]  \nonumber \\
    &\quad  %\left.
= \frac{1}{\eta} I,
\end{align}
with the parameters identified as
\begin{align}
a_1 &= 1 + \frac{B^2}{\eta^2} - n\sigma^2 \xi^2, \quad a_2 = n d_n \xi f, \quad a_3 = \frac{B}{\eta^2}, \quad a_4 = \frac{1}{\eta^2} - n\kappa^2,  \\
a_5 &= i\kappa^2 \sum_a l^a, \quad a_6 = -\frac{\kappa^2}{2}\sum_a (l^a)^2, \quad l^a \equiv d_n f - g^a, \quad d_n \equiv \frac{\sigma^2}{\sigma^2 + n(1+\sigma_0^2)},
\end{align}
where $\kappa^2 \equiv 1/(\sigma^2 + 1 - q)$. Since $\varphi(t, u) = 1$, we obtain
\begin{align}
e^{\widetilde{G}_{3,\mathrm{RS}}} &= \mathrm{Tr}_f \mathrm{Tr}_{\{g^a\}} \frac{\sigma^{n+1}}{\sqrt{\sigma^2 + n(1+\sigma_0^2)}} \widehat{a}^{-n/2} \frac{1}{\eta} I e^{G_{4,\mathrm{RS}}} \nonumber \\
&= \mathrm{Tr}_f \mathrm{Tr}_{\{g^a\}} \exp\left[ \frac{(a_1 a_5 + a_2 a_3)^2}{2 a_1(a_1 a_4 - a_3^2)} + \frac{a_2^2}{2 a_1} + a_6 + \ln D_n + G_{4,\mathrm{RS}} \right], \\
D_n &\equiv \frac{\sigma^{n+1}\widehat{a}^{-n/2}}{\eta \sqrt{\sigma^2 + n(1+\sigma_0^2)}\sqrt{a_1 a_4 - a_3^2}}.
\end{align}
Evaluating the terms in Eq.~\eqref{eq:formula} up to order  $\mathcal{O}(n)$ yields
\begin{align}
\frac{(a_1 a_5 + a_2 a_3)^2}{2 a_1(a_1 a_4 - a_3^2)} &= -n \frac{(1 - Q)(2m - q - (1+\sigma_0^2))}{2(\sigma^2 + 1 - q)^2} + {\cal O}(n^2), \\
a_2^2 &= {\cal O}(n^2), \\
\sum_a (l^a)^2 &= n d_n^2 + n - 2d_n \sum_a f g^a = 2n(1 - R) + {\cal O}(n^2), \\
\ln D_n &= n\ln\sigma - \frac{n}{2}\ln(\sigma^2 + 1 - q) + \frac{n}{2}\frac{2m - q - (1+\sigma_0^2)}{\sigma^2 + 1 - q} + {\cal O}(n^2).
\end{align}
%where $fg^a$ has been replaced by $R$.
Since we introduced $\delta$-functions such as Eq.~\eqref{eq:delta-function},
$f g^a$ and $g^a g^b$ have been replaced by $R$ and $Q$, respectively, in the equations above.
Therefore,  only $G_{4,\mathrm{RS}}$ depends on $f$ and $\{g^a\}$. Thus, $e^{G_{4,\mathrm{RS}}}$ becomes
\begin{align}
e^{G_{4,\mathrm{RS}}} &= \exp\left[ -\left(\widehat{r}_0 r_0 + n\widehat{r} r + n\widehat{R} R + \frac{n(n-1)}{2}\widehat{Q} Q\right) + \frac{1}{2}(\alpha_0 r_0^2 + \alpha n r^2) \right. \nonumber \\
&\quad \left. + \widehat{r}_0 f + \widehat{r}\sum_a g^a + \widehat{R} f \sum_a g^a + \widehat{Q}\sum_{a<b} g^a g^b \right].
\end{align}
%%%
Applying the H-S transformation to the quadratic term,
\begin{equation}
\exp\left[\widehat{Q}\sum_{a<b} g^a g^b\right] = \int Ds \exp\left[ \sqrt{\widehat{Q}} s \sum_a g^a - \frac{\widehat{Q}}{2}n \right],
\end{equation}
$e^{G_{4,\mathrm{RS}}}$ is simplified to
\begin{align}
e^{G_{4,\mathrm{RS}}} &= \exp\left[ -\left(\widehat{r}_0 r_0 + n\widehat{r} r + n\widehat{R} R - \frac{n}{2}\widehat{Q} Q\right) + \frac{1}{2}(\alpha_0 r_0^2 + \alpha n r^2) + \widehat{r}_0 f - \frac{\widehat{Q}}{2}n + {\cal O}(n^2) \right] \nonumber \\
&\quad \times \int Ds \exp\left[\Xi(s,f)\sum_a g^a\right], \label{eq:g4rs}
\end{align}
where $\Xi(s,f) \equiv \widehat{r} + \widehat{R}f + \sqrt{\widehat{Q}}s$.
Performing the trace over $\{g^a\}$ and $f$, we have
\begin{align}
\mathrm{Tr}_f \mathrm{Tr}_{\{g^a\}} e^{G_{4,\mathrm{RS}}} &= \exp\left[ -\left(\widehat{r}_0 r_0 + n\widehat{r} r + n\widehat{R} R - \frac{n}{2}\widehat{Q} Q\right) + \frac{1}{2}(\alpha_0 r_0^2 + \alpha n r^2) - \frac{\widehat{Q}}{2}n + {\cal O}(n^2) \right] \nonumber \\
&\quad \times \mathrm{Tr}_f e^{\widehat{r}_0 f} \int Ds \left( 2\cosh \Xi(s,f) \right)^n.
\end{align}
%%%
Expanding the last factor up to $O(n)$, we obtain
\begin{align}
\lefteqn{\mathrm{Tr}_f e^{\widehat{r}_0 f} \int Ds \left( 2\cosh \Xi(s,f) \right)^n} \nonumber \\
&= \mathrm{Tr}_f e^{\widehat{r}_0 f} \int Ds \left[ 1 + n\ln(2\cosh\Xi(s,f)) + {\cal O}(n^2) \right] \nonumber \\
&= \exp\left[ \ln(2\cosh\widehat{r}_0) + \frac{n}{2\cosh\widehat{r}_0} \mathrm{Tr}_f e^{\widehat{r}_0 f} \int Ds \ln(2\cosh\Xi(s,f)) + {\cal O}(n^2) \right].
\end{align}
Thus, up to order  $\mathcal{O}(n)$, the expression for $\mathrm{Tr}_f \mathrm{Tr}_{\{g^a\}} e^{G_{4,\mathrm{RS}}}$ becomes
\begin{align}
\mathrm{Tr}_f \mathrm{Tr}_{\{g^a\}} e^{G_{4,\mathrm{RS}}} &= \exp\left[ -\widehat{r}_0 r_0 + \frac{1}{2}\alpha_0 r_0^2 + \ln(2\cosh\widehat{r}_0) + n\left( -\widehat{r} r - \widehat{R} R - \frac{1}{2}\widehat{Q}(1-Q) \right.\right. \nonumber \\
&\quad \left.\left. + \frac{1}{2}\alpha r^2 + \frac{1}{2\cosh\widehat{r}_0}\mathrm{Tr}_f e^{\widehat{r}_0 f} \int Ds \ln(2\cosh\Xi(s,f)) \right) \right].
\end{align}
Consequently, $e^{\widetilde{G}_{3,\mathrm{RS}}}$ is obtained as
\begin{align}
e^{\widetilde{G}_{3,\mathrm{RS}}} &= \exp\left[ -\widehat{r}_0 r_0 + \frac{1}{2}\alpha_0 r_0^2 + \ln(2\cosh\widehat{r}_0) \right. \nonumber \\
&\quad + n\left( -\frac{(1-Q)(2m-q-(1+\sigma_0^2))}{2(\sigma^2+1-q)^2} - \frac{1-R}{\sigma^2+1-q} - \frac{1}{2}\ln\left(\frac{\sigma^2+1-q}{\sigma^2}\right) \right. \nonumber \\
&\quad + \frac{1}{2}\frac{2m-q-(1+\sigma_0^2)}{\sigma^2+1-q} - \widehat{r} r - \widehat{R} R - \frac{1}{2}\widehat{Q}(1-Q) + \frac{1}{2}\alpha r^2 \nonumber \\
&\quad \left.\left. + \frac{1}{2\cosh\widehat{r}_0}\mathrm{Tr}_f e^{\widehat{r}_0 f}\int Ds \ln(2\cosh\Xi(s,f)) \right) \right].
\end{align}
Decomposing $\widetilde{G}_{3,\mathrm{RS}}$  up to order $\mathcal{O}(n)$ as $\widetilde{G}_{3,\mathrm{RS}} = \widetilde{G}_{3,\mathrm{RS}}^0 + n\widetilde{G}_{3,\mathrm{RS}}^1$, we have\\
\begin{align}
\widetilde{G}_{3,\mathrm{RS}}^0 &= -\widehat{r}_0 r_0 + \frac{1}{2}\alpha_0 r_0^2 + \ln(2\cosh\widehat{r}_0), \\
\widetilde{G}_{3,\mathrm{RS}}^1 &= -\frac{(1-Q)(2m-q-(1+\sigma_0^2))}{2(\sigma^2+1-q)^2} - \frac{1-R}{\sigma^2+1-q} - \widehat{r} r - \widehat{R} R - \frac{1}{2}\widehat{Q}(1-Q) + \frac{1}{2}\alpha r^2 \nonumber \\
&\quad - \frac{1}{2}\ln\left(\frac{\sigma^2+1-q}{\sigma^2}\right) + \frac{1}{2}\frac{2m-q-(1+\sigma_0^2)}{\sigma^2+1-q} \nonumber \\
&\quad + \frac{1}{2\cosh\widehat{r}_0}\mathrm{Tr}_f e^{\widehat{r}_0 f} \int Ds \ln[2\cosh\Xi(s,f)]. \label{eq:rsnto0}
\end{align}
Note that $G_{3,\mathrm{RS}} = \widetilde{G}_{3,\mathrm{RS}}$ holds at the saddle point.
When $\alpha_0 > 1$, evaluating $Z_f$ at a non-zero saddle point yields the relation $\frac{1}{N}\ln Z_f = -\frac{1}{2}\alpha_0 r_0^2 + \ln(2\cosh\widehat{r}_0)$ for $N \gg 1$. Thus, we confirm that
\begin{equation}
G_{\mathrm{RS}}^0 = G_{2,\mathrm{RS}}^0 + \widetilde{G}_{3,\mathrm{RS}}^0 - \beta\ln 2 - \frac{1}{N}\ln Z_f = 0.
\end{equation}
Finally, the free energy per pixel, $\frac{G_{\mathrm{RS}}}{n} = G_{\mathrm{RS}}^1$, is derived as 
\begin{align}
G_{\mathrm{RS}}^1 &= G_{1,\mathrm{RS}}^1 + G_{2,\mathrm{RS}}^1 + G_{3,\mathrm{RS}}^1 \nonumber \\
&= \frac{1}{2}\beta q \widehat{q} - \beta m \widehat{m} - \frac{1}{2}\beta \widehat{q} + \beta \int Dz \ln \left[ 2\cosh\left(\sqrt{\widehat{q}} z + \widehat{m}\right) \right] \nonumber \\
&\quad - \frac{(1-Q)(2m-q-(1+\sigma_0^2))}{2(\sigma^2+1-q)^2} - \frac{1-R}{\sigma^2+1-q} - \widehat{r} r - \widehat{R} R - \frac{1}{2}\widehat{Q}(1-Q) + \frac{1}{2}\alpha r^2 \nonumber \\
&\quad - \frac{1}{2}\ln\left(\frac{\sigma^2+1-q}{\sigma^2}\right) + \frac{1}{2}\frac{2m-q-(1+\sigma_0^2)}{\sigma^2+1-q} \nonumber \\
&\quad + \frac{1}{2\cosh\widehat{r}_0}\mathrm{Tr}_f e^{\widehat{r}_0 f}\int Ds \ln \left[ 2\cosh\Xi(s,f) \right].
\end{align}
The corresponding SPEs are presented in Section~\ref{sec:formulation}.

%%%%%%%%%%%%%%%%%%%%%%%%%%%%%%%%%%%%%%%%%%%%%
\section{Calculation of the components of the Hessian matrix}
\label{sec:app_at_stability}

The AT stability condition is determined by the signs of eigenvalues for the Hessian matrix $\calH_{\mathrm{rp}}$ corresponding to the replicon mode --- specifically, the components involving $q_{ab}, \widehat{q}_{ab}, Q_{ab}$, and $\widehat{Q}_{ab}$.
The  Hessian matrix of $G$ is $n$ times that of the free energy $F$, and  the signs of their eigenvalues are identical. Therefore, we evaluate the Hessian matrix of $G$. Here, $G$ is given by $G = G_1 + G_2 + G_3$, where
\begin{align}
G_1 &= -\beta \sum_{a<b} \widehat{q}_{ab} q_{ab} - \beta \sum_a \widehat{m}_a m_a, \\
G_2 &= \beta \ln \left( \mathrm{Tr}_s \mathrm{Tr}_{\{x^a\}} e^{H_2} \right), \\
H_2 &\equiv \sum_{a<b} \widehat{q}_{ab} x^a x^b + \sum_a \widehat{m}_a s x^a, \\
Z_2 &\equiv \mathrm{Tr}_s \mathrm{Tr}_{\{x^a\}} e^{H_2}.
\end{align}
%%%
The non-zero second derivative of $G_1$ with respect to the order parameters is
\begin{equation}
\frac{\partial^2 G_1}{\partial q_{ab} \partial \widehat{q}_{cd}} = -\beta \delta_{(ab),(cd)}.
\end{equation}
%%%
Defining the expectation value $\langle A \rangle_2$ of a physical quantity $A$ as
\begin{equation}
\langle A \rangle_2 \equiv \frac{1}{Z_2} \mathrm{Tr}_s \mathrm{Tr}_{\{x^a\}} e^{H_2} A,
\end{equation}
the non-zero second derivative for $G_2$ is expressed as
\begin{equation}
\frac{\partial^2 G_2}{\partial \widehat{q}_{ab} \partial \widehat{q}_{cd}} = \beta \left( \langle x^a x^b x^c x^d \rangle_2 - \langle x^a x^b \rangle_2^2 \right).
\end{equation}
%%%
Under the RS ansatz, the corresponding Boltzmann factor and partition function reduce to
\begin{align}
e^{H_{2,\mathrm{RS}}} &= \int Dt \exp\left[ \left(\sqrt{\widehat{q}}t + \widehat{m}s\right) \sum_a x^a - \frac{n}{2}\widehat{q} \right] \nonumber \\
& \to \int Dt \exp\left[ \left(\sqrt{\widehat{q}}t + \widehat{m}s\right) \sum_a x^a \right] \quad (n \to 0), \\
Z_2 &= \mathrm{Tr}_s \int Dt e^{-\frac{n}{2}\widehat{q}} \left[ 2 \cosh\left(\sqrt{\widehat{q}}t + \widehat{m}s\right) \right]^n \to 2 \quad (n \to 0).
\end{align}
%%%
Consequently, the resulting expectation values yield
\begin{align}
\langle x^a s \rangle_{2,\mathrm{RS}} &= \int Dt \tanh\left(\sqrt{\widehat{q}}t + \widehat{m}\right) = m, \\
\langle x^a x^b \rangle_{2,\mathrm{RS}} &= \int Dt \tanh^2\left(\sqrt{\widehat{q}}t + \widehat{m}\right) = q \quad (a \neq b), \\
\langle x^a x^b x^c x^d \rangle_{2,\mathrm{RS}} &= \int Dt \tanh^4\left(\sqrt{\widehat{q}}t + \widehat{m}\right) \quad (\text{all indices } a, b, c, d \text{ distinct}).
\end{align}
Next, we evaluate the Hessian matrix of $G_3$. Since $\widetilde{G}_3 = G_3$ holds at the saddle point,
we use the explicit form of $\widetilde{G}_3$ and denote it by $G_3$.
The dependencies on the order parameters of ${\cal L}$ and $G_4$ are given by ${\cal L} = {\cal L}(\{m_a\}, \{q_{ab}\})$ and $G_4 = G_4(r_0, \widehat{r}_0, \{r_a\}, \{\widehat{r}_a\}, \{R_a\},
  \{\widehat{R}_a\}, \{Q_{ab}\}, \{\widehat{Q}_{ab}\})$.
We define the expectation value $\langle A \rangle_3$ as
\begin{align}
\langle A \rangle_3 &= \frac{1}{e^{G_3}} \mathrm{Tr}_f \mathrm{Tr}_{\{g^a\}} \left[ \prod_a \int \frac{\mathrm{d}\widehat{v}_a}{\sqrt{2\pi}} \right] \int Dt \, \exp[{\cal L} + G_4] A, \\
G_3 &= \ln \left( \mathrm{Tr}_f \mathrm{Tr}_{\{g^a\}} \left[ \prod_a \int \frac{\mathrm{d}\widehat{v}_a}{\sqrt{2\pi}} \right] \int Dt \, \exp[{\cal L} + G_4] \right).
\end{align}
%%%
The non-zero components of the Hessian matrix of $G_3$ are
\begin{align}
\frac{\partial^2 G_3}{\partial q_{ab} \partial q_{cd}} &= \langle \widehat{v}_a \widehat{v}_b \widehat{v}_c \widehat{v}_d \rangle_3 - \langle \widehat{v}_a \widehat{v}_b \rangle_3^2, \\
\frac{\partial^2 G_3}{\partial q_{ab} \partial \widehat{Q}_{cd}} &= -\langle \widehat{v}_a \widehat{v}_b g^c g^d \rangle_3 + \langle \widehat{v}_a \widehat{v}_b \rangle_3 \langle g^c g^d \rangle_3, \\
\frac{\partial^2 G_3}{\partial Q_{ab} \partial \widehat{Q}_{ab}} &= -1, \\
\frac{\partial^2 G_3}{\partial \widehat{Q}_{ab} \partial \widehat{Q}_{cd}} &= \langle g^a g^b g^c g^d \rangle_3 - \langle g^a g^b \rangle_3 \langle g^c g^d \rangle_3.
\end{align}
%%%
We now evaluate these expectations under the RS ansatz in the limit $n \to 0$.
Taking $n \to 0$, the definitions yield
\begin{equation}
\eta = \sqrt{2m - q}, \quad \widehat{b}(u, f, g^a) = u + if - ig^a, \quad B = i\xi\sigma^2,
\end{equation}
and $\exp[{\cal L}_{\mathrm{RS}}]$ simplifies to
\begin{equation}
\exp[{\cal L}_{\mathrm{RS}}] = \int Dz \exp\left[ -\frac{\widehat{a}}{2}\sum_a (\widehat{v}_a)^2 + \sum_a \widehat{b}(u, f, g^a) \widehat{v}_a \right].
\end{equation}
%%%
Taking $n \to 0$ in $e^{G_{4,\mathrm{RS}}}$, we have
\begin{equation}
e^{G_{4,\mathrm{RS}}} = \exp\left[ -\widehat{r}_0 r_0 + \widehat{r}_0 f + \frac{1}{2}\alpha_0 r_0^2 \right] \int Ds \exp\left[ \Xi(s,f)\sum_a g^a \right].
\end{equation}
The expectation value in the RS solution becomes
\begin{align}
\langle A \rangle_{3,\mathrm{RS}} &= \frac{1}{2\cosh\widehat{r}_0} \mathrm{Tr}_f e^{\widehat{r}_0 f} \mathrm{Tr}_{\{g^a\}} \left[ \prod_a \int \frac{\mathrm{d}\widehat{v}_a}{\sqrt{2\pi}} \right] \int Dt \int Dz \int Ds \nonumber \\
&\quad \times \exp\left[ -\frac{\widehat{a}}{2}\sum_a (\widehat{v}_a)^2 + \sum_a \widehat{b}(u, f, g^a)\widehat{v}_a \right] e^{\Xi(s,f)\sum_a g^a} A.
\end{align}
%%%
To evaluate the Hessian matrix, it is necessary to calculate the expectation values of $\widehat{v}_a\widehat{v}_b, \widehat{v}_a^2\widehat{v}_b^2, \widehat{v}_a^2\widehat{v}_b\widehat{v}_c$, and $\widehat{v}_a\widehat{v}_b\widehat{v}_c\widehat{v}_d$. 
For this purpose, we introduce the general $l$-body function $\calF_l^{(m)}(\{\widehat{v}_a\})$ of $m$ distinct variables, along with its counterpart obtained after integration $\psi_l^{(m)}(\widehat{a}, \{\widehat{b}_a\})$, defined by
\begin{eqnarray}
\lefteqn{\left[ \prod_a \int \frac{\mathrm{d}\widehat{v}_a}{\sqrt{2\pi}} \right]
    \exp\left[ -\frac{\widehat{a}}{2}\sum_a(\widehat{v}_a)^2
      + \sum_a\widehat{b}_a\widehat{v}_a \right] \calF_l^{(m)}(\{\widehat{v}_a\})} \nonumber \\ &=&
\left[ \prod_a 
      \frac{1}{\sqrt{\widehat{a}}}
      \exp\left[\frac{\widehat{b}_a ^2}{2\widehat{a}}\right] \right] \psi_l^{(m)}(\widehat{a}, \{\widehat{b}_a\}),
\end{eqnarray}
where $\widehat{b}_a=\widehat{b}(u,f,g^a)=u+  if -ig^a$.
Their explicit forms are given by
\begin{align}
\calF_2^{(2)}(\widehat{v}_a, \widehat{v}_b) &= \widehat{v}_a \widehat{v}_b, \quad \psi_2^{(2)}(\widehat{a}, \widehat{b}_a, \widehat{b}_b) = \frac{\widehat{b}_a}{\widehat{a}}\frac{\widehat{b}_b}{\widehat{a}}, \\
\calF_4^{(2)}(\widehat{v}_a, \widehat{v}_b) &= \widehat{v}_a^2 \widehat{v}_b^2, \quad \psi_4^{(2)}(\widehat{a}, \widehat{b}_a, \widehat{b}_b) = \left( \left(\frac{\widehat{b}_1}{\widehat{a}}\right)^2 + \frac{1}{\widehat{a}} \right) \left( \left(\frac{\widehat{b}_b}{\widehat{a}}\right)^2 + \frac{1}{\widehat{a}} \right),\\
\calF_4^{(3)}(\widehat{v}_a, \widehat{v}_b, \widehat{v}_c)&=\widehat{v}_a^2 \widehat{v}_b \widehat{v}_c, \quad \psi_4^{(3)}(\widehat{a}, \widehat{b}_a, \widehat{b}_b, \widehat{b}_c)
= \left( \left(\frac{\widehat{b}_a}{\widehat{a}}\right)^2 + \frac{1}{\widehat{a}} \right)  \frac{\widehat{b}_b}{\widehat{a}}\frac{\widehat{b}_c}{\widehat{a}},\\
\calF_4^{(4)}(\widehat{v}_a, \widehat{v}_b, \widehat{v}_c, \widehat{v}_d)
&=\widehat{v}_a \widehat{v}_b \widehat{v}_c\widehat{v}_d,
\quad \psi_4^{(4)}(\widehat{a}, \widehat{b}_a, \widehat{b}_b, \widehat{b}_c, \widehat{b}_d)
 =\frac{\widehat{b}_a}{\widehat{a}}   \frac{\widehat{b}_b}{\widehat{a}}\frac{\widehat{b}_c}{\widehat{a}}
\frac{\widehat{b}_d}{\widehat{a}}.
\end{align}
Thus, we obtain
\begin{align}
  \langle \calF_l^{(m)}(\{\widehat{v}_a\})  
  \rangle_{3,\mathrm{RS}} &=   \widehat{a}^{-n/2} \frac{1}{2\cosh\widehat{r}_0} \mathrm{Tr}_f e^{\widehat{r}_0 f} \mathrm{Tr}_{\{g^a\}} \int Dt \int Dz \int Ds \, e^{\Xi(s,f)\sum_a g^a} \nonumber \\
  &\quad \times \exp\left[ \frac{1}{2(\sigma^2 + 1 - q)} \sum_a (u + if - ig^a)^2 \right]\psi_l^{(m)}(\widehat{a}, \{u + if - ig^a\}).
\end{align}
Let us apply the Gaussian integration formula, Eq.~\eqref{eq:formula}.
In the limit $n \to 0$, the variable transformation $(t,u) \to (x_1, x_2)$ reduces to
\begin{align}
x_1 &= \sqrt{\frac{2m - q - (1+\sigma_0^2)}{2m - q}} \left( t - i\frac{\sqrt{1+\sigma_0^2}u}{2m - q - (1+\sigma_0^2)} \right), \\
x_2 &= \frac{1}{\sqrt{2m - q - (1+\sigma_0^2)}} \left( u - i\frac{2m - q - (1+\sigma_0^2)}{\sigma^2 + 1 - q} \sum_b g^b \right),
\end{align}
and we get
\begin{align}
  \langle   \calF_l^{(m)}(\{\widehat{v}_a\})  
  \rangle_{3,\mathrm{RS}} &=   \widehat{a}^{-n/2} \frac{1}{2\cosh\widehat{r}_0} \mathrm{Tr}_f e^{\widehat{r}_0 f} \mathrm{Tr}_{\{g^a\}} \int Dx_2 \int Ds \, e ^{\Xi(s,f)\sum_a g^a} \nonumber \\
  &\quad \times \psi_l^{(m)}(\widehat{a}, \{b_4x_2+b_5+ if - ig^a\}). 
\end{align}
Defining $\zeta \equiv \kappa^2(\eta^2 + B^2) = \frac{2m - q - (1+\sigma_0^2)}{\sigma^2 + 1 - q}$ and ${\cal S} \equiv \sum_b g^b$, in the limit $n \to 0$, we obtain
\begin{equation}
b_5  = i\kappa^2(\eta^2 + B^2)\sum_a(f - g^a)  \to -i\zeta{\cal S}.
\end{equation}
Therefore, in the limit $n \to 0$, we find
\begin{align}
  \langle   \calF_l^{(m)}(\{\widehat{v}_a\})  
  \rangle_{3,\mathrm{RS}} &= \frac{1}{2\cosh\widehat{r}_0} \mathrm{Tr}_f e^{\widehat{r}_0 f} \mathrm{Tr}_{\{g^a\}}  \int Dx_2 \int Ds \, e ^{\Xi(s,f)\sum_a g^a} \nonumber \\
  &\quad \times  \psi_l^{(m)}(\widehat{a}, \{b_4x_2+i(-\zeta{\calS} + f - g^a)\}),
\end{align}
where $b_4 \equiv \sqrt{2m - q - (1+\sigma_0^2)}$. 
Then, including a general  function  $\calG$ of $f$ and $\{g^a\}$, we obtain
\begin{align}
  \langle \calF_l^{(m)}(\{\widehat{v}_a\}) \calG(f,\{g^a\}) \rangle_{3,\mathrm{RS}} &= \frac{1}{2\cosh\widehat{r}_0}
  \mathrm{Tr}_f e^{\widehat{r}_0 f} \mathrm{Tr}_{\{g^a\}} \int Ds \, e^{\sum_a \Xi(s,f)g^a} \calG(f, \{g^a\}) \nonumber \\
  &\quad \times \int Dx_2 \, \psi_l^{(m)}\left(\kappa^{-2}, \{b_4 x_2 + i(-\zeta S +f -g^a)\}\right)
 \label{eq:avA}.
\end{align}
For instance, the expectation $\langle \widehat{v}_a \widehat{v}_b \rangle_{3,\mathrm{RS}}$ is given by
\begin{equation}
\langle \widehat{v}_a \widehat{v}_b \rangle_{3,\mathrm{RS}} = \frac{1}{(\sigma^2 + 1 - q)^2} \left[ \widehat{R}\{2m - q - (1+\sigma_0^2)\}(\sigma^2 - 1 - q + 2Q) - (Q - 2R + 1) \right].
\end{equation}
%%%
Furthermore, the average of $\calG(f, \{g^a\})$ reduces to
\begin{equation}
\langle \calG(f, \{g^a\}) \rangle_{3,\mathrm{RS}} = \frac{1}{2\cosh\widehat{r}_0} \mathrm{Tr}_f e^{\widehat{r}_0 f} \int Ds \, \mathrm{Tr}_{\{g^a\}} e^{\Xi(s,f)\sum_a g^a} \calG(f, \{g^a\}).
\end{equation}
%%%
For any function $h(\Xi(s,f))$, we define the shorthand notation
\begin{equation}
\overline{h f^a} \equiv \frac{1}{2\cosh\widehat{r}_0} \mathrm{Tr}_f e^{\widehat{r}_0 f}f^a \int Ds \, h(\Xi(s,f)),
\end{equation}
where $a=0$ or 1.
Setting $t_h \equiv \tanh(\Xi(s,f))$, the order parameters are compactly expressed as
\begin{equation}
r = \langle g^a \rangle_{3,\mathrm{RS}} = \overline{t_h}, \quad R = \langle f g^a \rangle_{3,\mathrm{RS}} = \overline{t_h f}, \quad Q = \langle g^a g^b \rangle_{3,\mathrm{RS}} = \overline{t_h^2}.
\end{equation}
Focusing on the replicon sector, the replicon Hessian matrix $\calH_{\mathrm{rp}}$ of size $4n_1 \times 4n_1$ (where $n_1 \equiv \binom{n}{2} = \frac{n(n-1)}{2}$) is formulated as
\begin{equation}
\calH_{\mathrm{rp}} = 
\begin{pmatrix}
\left( \frac{\partial^2 G}{\partial q_{ab} \partial q_{cd}} \right) & \left( \frac{\partial^2 G}{\partial q_{ab} \partial \widehat{q}_{cd}} \right) & \left( \frac{\partial^2 G}{\partial q_{ab} \partial Q_{cd}} \right) & \left( \frac{\partial^2 G}{\partial q_{ab} \partial \widehat{Q}_{cd}} \right) \\[1ex]
\left( \frac{\partial^2 G}{\partial \widehat{q}_{ab} \partial q_{cd}} \right) & \left( \frac{\partial^2 G}{\partial \widehat{q}_{ab} \partial \widehat{q}_{cd}} \right) & \left( \frac{\partial^2 G}{\partial \widehat{q}_{ab} \partial Q_{cd}} \right) & \left( \frac{\partial^2 G}{\partial \widehat{q}_{ab} \partial \widehat{Q}_{cd}} \right) \\[1ex]
\left( \frac{\partial^2 G}{\partial Q_{ab} \partial q_{cd}} \right) & \left( \frac{\partial^2 G}{\partial Q_{ab} \partial \widehat{q}_{cd}} \right) & \left( \frac{\partial^2 G}{\partial Q_{ab} \partial Q_{cd}} \right) & \left( \frac{\partial^2 G}{\partial Q_{ab} \partial \widehat{Q}_{cd}} \right) \\[1ex]
\left( \frac{\partial^2 G}{\partial \widehat{Q}_{ab} \partial q_{cd}} \right) & \left( \frac{\partial^2 G}{\partial \widehat{Q}_{ab} \partial \widehat{q}_{cd}} \right) & \left( \frac{\partial^2 G}{\partial \widehat{Q}_{ab} \partial Q_{cd}} \right) & \left( \frac{\partial^2 G}{\partial \widehat{Q}_{ab} \partial \widehat{Q}_{cd}} \right)
\end{pmatrix}.
\end{equation}
%%%
Here, each sub-block consists of the second-order partial derivatives of  $G$ with respect to the order parameters $\{q_{ab}, \widehat{q}_{ab}, Q_{ab}, \widehat{Q}_{ab}\}$. 
For instance, the sub-block $\left( \frac{\partial^2 G}{\partial q_{ab} \partial q_{cd}} \right)$ represents an $n_1 \times n_1$ matrix whose $((ab), (cd))$-component is given by $\frac{\partial^2 G}{\partial q_{ab} \partial q_{cd}}$ with $a<b$ and $c<d$.
Explicitly, $\calH_{\mathrm{rp}}$ is expressed as
\begin{equation}
\calH_{\mathrm{rp}} = 
\begin{pmatrix}
H(P_3, Q_3, R_3) & -\beta E_{n_1} & 0_{n_1} & H(\widetilde{P}_3, \widetilde{Q}_3, \widetilde{R}_3) \\
-\beta E_{n_1} & H(P_3', Q_3', R_3') & 0_{n_1} & 0_{n_1} \\
0_{n_1} & 0_{n_1} & 0_{n_1} & -E_{n_1} \\
H(\widetilde{P}_3, \widetilde{Q}_3, \widetilde{R}_3) & 0_{n_1} & -E_{n_1} & H(\widehat{P}_3, \widehat{Q}_3, \widehat{R}_3)
\end{pmatrix},
\end{equation}
where $E_{n_1}$ and $0_{n_1}$ denote the $n_1 \times n_1$ identity and zero matrices, respectively, 
and  $H(P, Q, R)$ represents the standard $n_1 \times n_1$ replicon matrix defined by
\begin{equation}
H(P,Q, R)_{(ab)(cd)} = \begin{cases}
P & \text{if } (cd) = (ab), \\
Q & \text{if } c \in \{a,b\} \text{ or } d \in \{a,b\} \text{ (with } (cd) \neq (ab)\text{)}, \\
R & \text{if } \{a,b\} \cap \{c,d\} = \emptyset.\\
\end{cases}
\end{equation}
The matrix elements $P_3, Q_3, R_3$, etc., are evaluated as
\begin{align}
P_3 &= \langle (\widehat{v}_a)^2 (\widehat{v}_b)^2 \rangle_{3,\mathrm{RS}} - \langle \widehat{v}_a \widehat{v}_b \rangle_{3,\mathrm{RS}}^2, \\
Q_3 &= \langle (\widehat{v}_a)^2 \widehat{v}_b \widehat{v}_c \rangle_{3,\mathrm{RS}} - \langle \widehat{v}_a \widehat{v}_b \rangle_{3,\mathrm{RS}}^2, \\
R_3 &= \langle \widehat{v}_a \widehat{v}_b \widehat{v}_c \widehat{v}_d \rangle_{3,\mathrm{RS}} - \langle \widehat{v}_a \widehat{v}_b \rangle_{3,\mathrm{RS}}^2, \\
\widetilde{P}_3 &= -\langle \widehat{v}_a \widehat{v}_b g^a g^b \rangle_{3,\mathrm{RS}} + \langle \widehat{v}_a \widehat{v}_b \rangle_{3,\mathrm{RS}} Q, \\
\widetilde{Q}_3 &= -\langle \widehat{v}_a \widehat{v}_b g^a g^c \rangle_{3,\mathrm{RS}} + \langle \widehat{v}_a \widehat{v}_b \rangle_{3,\mathrm{RS}} Q, \\
\widetilde{R}_3 &= -\langle \widehat{v}_a \widehat{v}_b g^c g^d \rangle_{3,\mathrm{RS}} + \langle \widehat{v}_a \widehat{v}_b \rangle_{3,\mathrm{RS}} Q, \\
P_3' &= \beta (1 - q^2), \quad Q_3' = \beta q(1 - q), \quad R_3' = \beta (\overline{r} - q^2), \\
\widehat{P}_3 &= 1 - Q^2, \quad \widehat{Q}_3 = Q(1 - Q), \quad \widehat{R}_3 = \overline{R} - Q^2,
\end{align}
where we define
\begin{align}
\overline{R} &\equiv \langle g^a g^b g^c g^d \rangle_{3,\mathrm{RS}} = \overline{t_h^4}, \\
\overline{r} &\equiv \langle x^a x^b x^c x^d \rangle_{2,\mathrm{RS}} = \int Dt \tanh^4\left(\sqrt{\widehat{q}}t + \widehat{m}\right).
\end{align}

%%%%%%%%%%%%%%%%%%%%%%%%%%%%%%%%%%%%%%%%%%%     

\end{document}